\documentclass[%
superscriptaddress,
preprint,
amsmath,amssymb,
aps,
prx,
floatfix,
]{revtex4-2}

\usepackage{graphicx}% Include figure files
\usepackage{dcolumn}% Align table columns on decimal point
\usepackage{bm}% bold math
\usepackage{hyperref}% add hypertext capabilities
\usepackage[mathlines]{lineno}% Enable numbering of text and display math
\usepackage{amsmath}
\usepackage{physics}
\usepackage{caption}
\usepackage{mathrsfs}

\begin{document}

%\preprint{APS/123-QED}

\title{A Long-distance Quantum-capable Internet Testbed}% Force line breaks with \\

\author{Dounan Du}
\affiliation{Department of Physics and Astronomy, Stony Brook University, Stony Brook, NY 11794, USA}
\author{Leonardo Castillo-Veneros}
\affiliation{Department of Physics and Astronomy, Stony Brook University, Stony Brook, NY 11794, USA}
\author{Dillion Cottrill}
\affiliation{Department of Physics and Astronomy, Stony Brook University, Stony Brook, NY 11794, USA}
\author{Guo-Dong Cui}
\affiliation{Department of Physics and Astronomy, Stony Brook University, Stony Brook, NY 11794, USA}
%\author{Gabriel Bello}
%\affiliation{Qunnect Inc., 141 Flushing Av. Suite 1110, Brooklyn, NY, 11205, USA}
%\author{Mael Flament}
%\affiliation{Qunnect Inc., 141 Flushing Av. Suite 1110, Brooklyn, NY, 11205, USA}
\author{Paul Stankus}
\affiliation{Brookhaven National Laboratory, Upton, NY, 11973, USA}
\author{Dimitrios Katramatos}
\affiliation{Brookhaven National Laboratory, Upton, NY, 11973, USA}
\author{Juli\'an Mart\'inez-Rinc\'on}
\affiliation{Brookhaven National Laboratory, Upton, NY, 11973, USA}
\author{Eden Figueroa}
 \email{Contact author: eden.figueroa@stonybrook.edu}
\affiliation{Department of Physics and Astronomy, Stony Brook University, Stony Brook, NY 11794, USA}
\affiliation{Brookhaven National Laboratory, Upton, NY, 11973, USA}


\begin{abstract}
Building a Quantum Internet requires the development of innovative quantum-enabling networking architectures that integrate advanced communication systems with long-distance quantum communication hardware. Here, we present the implementation of a quantum-enabled internet prototype using a novel physics-centric stack-based quantum network paradigm to govern the dynamics of multiple light-matter Hamiltonians across distant nodes. We demonstrate this concept using a deployed large-scale intercity quantum network connecting laboratories at Stony Brook University and the Brookhaven National Laboratory. This network facilitates a fundamental long-distance quantum network service—that of high-visibility Hong-Ou-Mandel interference of telecom quantum states generated in two independent, telecom-compatible quantum light-matter interfaces separated by a distance of 158 km.
%\begin{description}
%use the optional argument of the \verb+\item+ command to give %the category of each item. 
%\end{description}
\end{abstract}

\maketitle

%\tableofcontents

\section{\label{sec:1}Introduction}

Entanglement-based Quantum Networks (QNs) have great potential to enhance information processing, secure communication, and bolster fundamental scientific research\cite{osti_2001045}. Targeting this promise, the pioneering concept of a quantum internet was developed \cite{Kimble2008}, in which dynamic light-matter interconnects are linked using point-to-point quantum communication channels. Fundamental capabilities of this early concept, such as direct entanglement distribution\cite{Hensen2016, Valivarthi2016, Sun2016, Valivarthi2020}, quantum-state transfer\cite{Puigibert2020, Cao_2020}, and interference-mediated entanglement generation between light-matter quantum nodes\cite{Slodicka2013}, have been demonstrated recently\cite{vanLeent2022, LagoRivera2021, Yu2020, Neumann2022}. Furthermore, QNs designed towards large-scale Distributed Quantum Processing present a promising roadmap to achieving quantum advantage\cite{OhioWorskhop}.  

Advancing these pioneering concepts towards demonstrating a large-scale Quantum Internet (QI)\cite{Wehner2018} is a formidable challenge. A fundamental step is to demonstrate first-generation measurement-based entanglement distribution protocols\cite{azuma2023quantum,muralidharan2016optimal}, such as Cabrillo entanglement generation\cite{cabrillo1999creation}, Duan-Lukin-Cirac-Zoller (DLCZ) entanglement creation, and memory-assisted entanglement swapping\cite{PhysRevLett.80.3891} and teleportation\cite{bouwmeester1997experimental}, using deployed fiber and intercity quantum connections. The construction of such long-distance measurement-based QNs has several fundamental requirements: (i) synchronized control of multiple time-dependent Hamiltonians, governing the dynamics of light-matter interfaces located at remote nodes, (ii) maintaining consistent and robust qubit states across nodes, a task made arduous by fiber inhomogeneities and environmental fluctuations\cite{PhysRevApplied.21.014024}, (iii) achieving high-fidelity measurement-based QN operations, which are usually benchmarked through qubit indistinguishability via interference measurements such as Hong-Ou-Mandel (HOM)\cite{gera2024hong}, (iv) addressing the exponential attenuation of telecom channels with distance using memory-based quantum repeater systems\cite{RevModPhys.95.045006} and, (v) developing network-level control systems that take into account the inherently probabilistic nature of long-distance quantum network measurements\cite{yu2020entanglement}, considering QN operations as quantum processes in a large Hilbert space\cite{de2024quantum,karimi}. These last considerations diverge significantly from the deterministic operation of classical networks\cite{rfc8557,rfc9023}. 

From a computer science perspective, several groups have explored layered architectures to organize the quantum communication process by resource-dedicated tasks or protocols \cite{Pirker_2019}, analogous to the TCP/IP protocol stack for classical communication. First steps towards implementing these stack abstraction models have been demonstrated at the laboratory scale\cite{pompili}. However, a large-scale intercity quantum network based on a stack abstraction model governing the dynamics of multiple Hamiltonians across several nodes has not been demonstrated. Addressing this multifaceted challenge calls for innovative quantum-enabling networking architectures that integrate advanced communication systems, long-distance quantum communication networks, and advanced control systems that effectively manage and synchronize QN processes and communicate the outcome of probabilistic measurements that alter the QN status.
 
This work presents a new physics-centric stack-based QN paradigm using state-of-the-art classical and quantum communication systems. We implement this new methodology following contemporaneous networking principles while considering the unique systematic and physical demands of an Internet enabled to execute quantum operations across a collection of networked Hamiltonians: a Quantum-Enabled Internet (QEI). We use optical fiber-based networks, which are excellent candidates to build high-repetition-rate, long-distance QNs, offering a versatile platform synergizing with the established telecommunication infrastructure\cite{sun2017entanglement,neumann2022continuous}, enhancing cost-effectiveness. Additionally, we use time-evolving light-matter interconnects based on atomic systems with energy transitions at telecom wavelengths\cite{van2022entangling, van2020long,krutyanskiy2024multimode}, facilitating efficient quantum information transfer across QN nodes. Lastly, we develop and demonstrate the concept of a quantum-enabling architecture that merges advanced communication management principles and high-fidelity QN services.

\section{\label{sec:architecture}Quantum-Enabled Internet Paradigm}

\begin{figure*}
\includegraphics[width=1.0\textwidth]{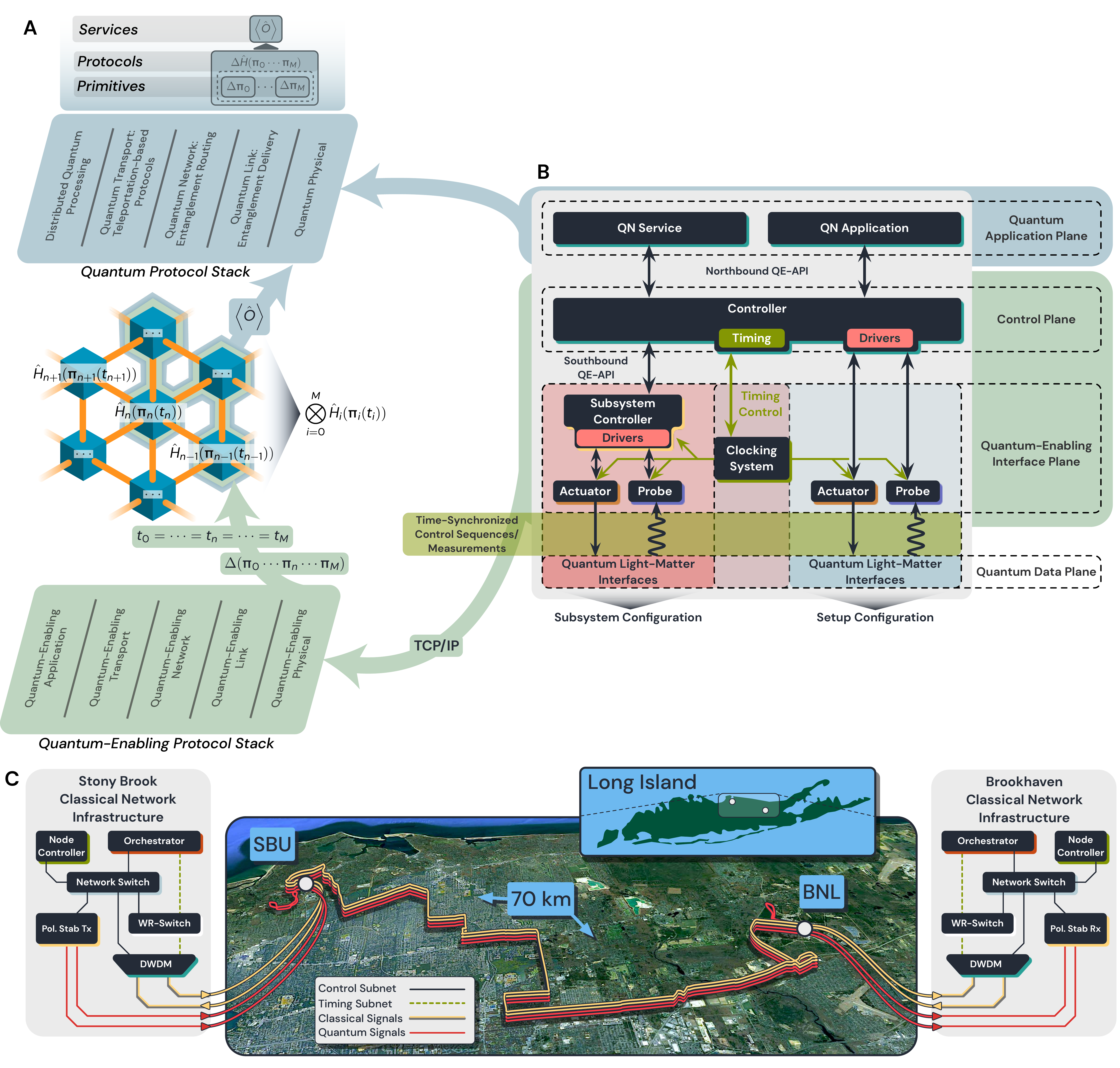}
\caption{\label{fig:1merged}(a) A quantum network consists of non-local Hamiltonians allowing hierarchical operations, with fundamental processes defined as primitives. These primitives—temporal modulations of Hamiltonian parameters—establish network protocols that yield services as outcomes. These protocols may be organized into a quantum protocol stack for quantum communication networks. Realizing this concept requires ancillary structures to enable robust construction of quantum primitives and protocols. These quantum-enabling processes can be grouped into a supporting protocol stack. Fundamental quantum-enabling protocols involve preparing and controlling physical parameters corresponding to Hamiltonian parameters and distributing a global sense of time throughout the network. (b) Our SDN-inspired quantum-enabling (QE) architecture abstracts the implementation of QE processes. QE processes are orchestrated by the control plane and executed at the quantum-enabling interface plane. Control thus trickles down to the quantum data plane, resulting in forwarding of quantum information. Control commands and measurements are exchanged between planes via quantum-enabling application programming interfaces. (c) SBU-BNL optical infrastructure uses four ~70 km optical fibers. One pair (red) is for quantum information, and another pair (yellow) is for time synchronization and classical data signals. The Alice, Bob, and Dave quantum nodes are co-located at SBU, with the Charlie node at BNL. Additional fiber loops at SBU and BNL allow phase independence in the quantum channels.}
\end{figure*}

To build a first instance of a long-distance measurement-based QEI, we propose a quantum network paradigm adopting an operational hierarchy correspondent to that of classical networks (see Fig.~\ref{fig:1merged}(a) top)\cite{OSI}. The heart of the abstraction is to consider QNs as a collection of dynamic Hamiltonians located at the QN nodes and the QN protocols as QN processes \cite{de2024quantum} obtained by simultaneously driving and measuring sets of remotely-located Hamiltonians (see Fig.~\ref{fig:1merged}(a) center). We break down the complex processes of a large QN into \textbf{QN primitives}—temporal manipulations of Hamiltonian parameters critical for QN nodes' operations. These primitives are grouped into \textbf{QN protocols} that guide the evolution of the entire network. Finally, key outcomes deriving from quantum observables associated with fundamental QN protocols (e.g., long-distance HOM interference or entanglement distribution and swapping) are defined as \textbf{QN services}.

To implement quantum protocols and develop quantum communication services and standards, we introduce the Quantum-Enabling (QE) Stack. The QE Stack consists of deterministic \textbf{QE-protocols} needed to prepare, control, and monitor quantum network operations at every step of the communication process (Fig.~\ref{fig:1merged}(a), green stack). QE protocols are responsible for physically realizing QN primitives and protocols. This establishes the guidelines for using ancillary classical networks to support QN services and to develop the Quantum Internet stack (Fig.~\ref{fig:1merged}(a), blue stack).

In this architecture, the execution of QE protocols and the operation of a QEI are understood through an architectural model inspired by Software-Defined Networking (SDN) concepts~\cite{KuroseRoss16, Simmons_2014}. It consists of four coexisting planes: a Quantum Application Plane (QAP), a Control Plane (CP), a Quantum-Enabling Interface Plane (QEIP), and a Quantum Data Plane (QDP)(Fig.~\ref{fig:1merged}(b)). We define this as the Quantum-enabling network architecture. Here, we follow the traditional networking notion of a plane, a telecommunications architecture representing specific components, processes, and types of traffic~\cite{Simmons_2014, rfc3654}. 

The QDP is responsible for manipulating and transporting quantum information (see Fig.~\ref{fig:1merged}(b) bottom). As control from the CP trickles down through the QEIP, the QDP enables the operation of quantum devices toward orchestrated actions, such as generating entangled photons, quantum memory operations, and quantum measurements. QN primitives and protocols are realized in the QDP. 

The QEIP is at the core of the QN operation, comprising all control and interfacing elements for the multitude of systems forming the QN. Control and interfacing elements are grouped by functionality into setups and subsystems. Elements within setup configurations are entirely managed by the central network controller. Within subsystem configurations, the central controller delegates control to task-specific controllers (see Fig.~\ref{fig:1merged}(b) middle). Additionally, to maintain temporal accordance across the entire network, the QEIP requires a precise clocking interface to distribute timing information to setups and subsystems. Our QEIP utilizes a White-Rabbit (WR) network\cite{Lipinski_2011} to establish network clocking with a temporal jitter shorter than the temporal duration of the photons propagating through the QN. 

The CP is centralized and responsible for managing and forwarding quantum information by executing QE protocols through the QEIP. Southbound QE Application Programming Interfaces (APIs) enable communication from the CP to the QEIP, allowing for non-local entanglement-based operations. Northbound QE-APIs allow users or higher-level functions to interface with the central controller and obtain a network-wide view of available capabilities.

The QAP provides the facade for all network operations, allowing them to be addressed independently of lower-level operations. The QAP encompasses higher-level entanglement-based applications, such as distributed quantum gate operation and teleportation-based services, and lower-level QN services, such as HOM indistinguishability verification or entanglement swapping (see Fig.~\ref{fig:1merged}(b) top). This enables user-defined network operation, interaction between QN protocol layers, and the development of more complicated network operations.

\section{\label{sec:implementation}Long Island Quantum-Enabled Internet Prototype Implementation}

Our QEI paradigm is implemented using the infrastructure of the New York State Quantum Internet Testbed (NYSQIT) located across Long Island. This deployed large-scale QN system facilitates the execution of fundamental quantum services and protocols described above, with a vision to support entanglement distribution and quantum state teleportation across extended distances. We leverage the telecommunication fiber infrastructure connecting Stony Brook University (SBU) and Brookhaven National Laboratory (BNL), consisting of four commercial single-mode telecommunication fibers, each spanning approximately 70 km (see Fig.~\ref{fig:1merged}(c)). In the following, we describe our developments to implement a quintessential QEI service: a long-distance high-visibility HOM interference of telecom photonic qubits produced in two independent light-matter systems connected by 158 kilometers of fiber using our QEI network architecture. 

\subsection{\label{sec:3.2}Quantum Data Plane Implementation}
For our QDP implementation, we have chosen quantum systems capable of high-fidelity production of quantum states and transfer of telecom-compatible quantum states, including (i) field-deployable quantum-memory-compatible quantum frequency converters (QM-QFC) operating at telecom wavelengths and (ii) HOM interference stations used to evaluate the interoperability among the QM-QFCs and the quantum interference visibility of qubits after transmission over long distances. These systems are distributed among four nodes located at SBU and BNL.

\subsubsection{Light-matter Quantum Interfaces: Quantum-Memory-Compatible Quantum Frequency Converters}

\begin{figure*}
\centering
\includegraphics[width=0.9\textwidth]{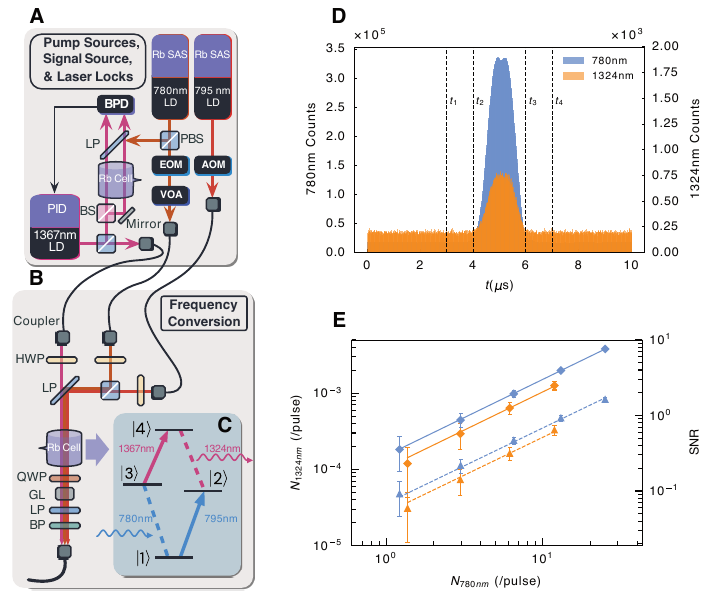}
\caption{\label{fig:2merged}Experimental QM-QFC setup. (a) Frequency locking of the $\rm 1367nm$ laser using an Optical-Optical Double Resonance(OODR) laser stabilization setup (see Methods for details). (b) An EOM creates the field envelope of $\rm 780nm$ signal. The attenuated $\rm 780nm$ signal is combined with the $\rm 795nm$ and the $\rm 1367nm$ pump light. After the cell the pump beams are filtered out by a GL polarizer and wavelength filters. (c) The atomic diamond-shaped scheme used for frequency conversion. (d) Input and output pulses detected by the SNSPDs used to estimate the efficiency of the QFC process. In the cycle time of $10\mu s$, $t_1 = 3\mu s$, $t_2 = 4\mu s$, $t_3 = 6\mu s$ and $t_4 = 7\mu s$. Photon numbers are evaluated within $[t_2, t_3]$ interval and the backgroud noise photon rate is evaluated from $[0 \mu s, t_1]$ and $[t_4, 10\mu s]$. (e) Measured number of photons of QFC pulses at 1324 nm (diamonds) and signal-to-noise ratio (SNR) at the detector (triangles) as a function of the mean photon number in the 780 nm input pulse for Alice (orange) and Bob (blue). The solid and dash lines are linear fittings to the conversion efficiency and SNR curves, respectively. The data are generated from calculating mean values after $\sim$20 minutes of interrogation time for pulse repetition rates of 100 kHz. The error bars are dominated by detection and conversion efficiency errors. LD, Laser diode; PBS, polarizing beam splitter; SAS, saturated absorption spectroscopy; QWP/HWP, quarter/half-wave plate; BP/LP, bandpass/longpass filters; GL, Glan laser polarizer.}
\end{figure*}

At SBU, we field two identical QM-QFCs. Based on rubidium atomic vapor, these devices are derived from our room-temperature quantum memory platform\cite{gera2024hong}. Fig.~\ref{fig:2merged}(c) demonstrates the four-level diamond shape energy scheme used for the QFC process. The input $\rm 780 ~nm$ weak coherent state simulates the readout qubits from quantum memories. The input field resonantly couples the ground state $\ket{1}$($\rm 5S_{1/2}, F=3$) to the excited state $\ket{3}$($\rm 5P_{3/2},F=4$). The $\rm795 ~nm$ pump I couples the ground state $\ket{1}$ to the excited state $\ket{2}$($\rm 5P_{1/2},F=3$), and the $\rm 1367 ~nm$ pump II couples $\ket{3}$ to a higher excited state $\ket{4}$($\rm 6S_{1/2},F=3$). The conversion system is governed by the Four-Wave Mixing (FWM) Hamiltonian (see SM for derivation details):
\begin{widetext}
\begin{equation}
    \begin{aligned}
    \tilde{H}_{FWM}&=\hbar\Delta_a \hat{a}^{\dagger}\hat{a}+\hbar\Delta_b\hat{b}^{\dagger}\hat{b}
    +\sum^N_{i=1}\left[\hbar\Delta_2\sigma^i_{22}+\hbar\Delta_3\sigma^i_{33}+\Delta_4\sigma^i_{44}+\right.\\
    &\left.\left(\hbar\frac{\Omega_\text{I}}{2}\sigma^i_{21}+\hbar\frac{\Omega_\text{II}}{2}\sigma^i_{43}
    +\hbar g_{p}\hat{a}\sigma^i_{31}+\hbar g_{QFC}\hat{b}\sigma^i_{42}+h.c.\right)\right]
    \end{aligned}
\end{equation}
\end{widetext}

where $N$ is the number of atoms in the ensemble, $\hat{a}^\dagger$ is the creation operator of the input probe field, $\hat{b}^\dagger$ is the creation operator for the QFC output field connecting states $\ket{2}$ and $\ket{4}$, $\Omega_I$ and $\Omega_{II}$ are the Rabi frequencies of the two pumps fields, $g_p$ and $g_{QFC}$ are the dipole-coupling strengths for the input and the output quantum fields, $\sigma_{jk}$ is the $jk$th element of the four-dimensional atomic operator. The $\Delta_i$'s are the single photon detunings. By the input-output formalism, one can show the frequency conversion from the $\rm 780 nm$ input field to the $\rm 1324nm$ output field follows the linear relation$\langle\hat{b}_{out}^{\dagger}(t)\hat{b}_{out}(t)\rangle=|\eta(0)|^2\langle\hat{a}_{in}^{\dagger}(t)\hat{a}_{in}(t)\rangle$ where $|\eta(0)|^2$ is a constant conversion efficiency independent of the mean input photon number(see SM for the complete derivation). Both rubidium vapor cells contain naturally abundant $\rm ^{85}Rb$ and $\rm ^{87}Rb$ and are magnetically shielded. The setup used on both systems is shown in Fig.~\ref{fig:2merged}(a) and (b). For each of the QFC, the modification of the set of parameters of the system $\pi_\text{Alice/Bob}=\{\Omega_\text{I}(t),\ \Omega_\text{II}(t),\ \langle \hat{a}^{\dagger}(t)\hat{a}(t)\rangle \}$ constitutes the QN primitives. 

The FWM QFC process can be driven with efficiencies higher than 50\% \cite{kumar2023quantum}. In our experiments, we achieve a compromise by lowering the QFC efficiency and achieving an operational point with a good signal-to-noise ratio (SNR), which is the central figure of merit for HOM interference experiments. To measure the conversion efficiencies, we use streams of $\sim0.9\mu s$ FWHM pulses, with Alice's Pump I operating at 5.20 mW and Pump II at 2.95 mW, while Bob's Pump I is set to 0.45 mW and Pump II to 2.20 mW.  The input pulses are detected using Superconducting Nanowire Single-Photon Detectors (SNSPD), and mean photon numbers for the probe and FWM QFC converted fields were estimated using the resultant histograms (see Fig.~\ref{fig:2merged}(d), 780-blue and 1324-orange). We then characterized the conversion efficiencies and signal-to-noise ratios for different input photon levels. The conversion efficiencies for the two QM-QFCs (Alice and Bob) are measured to be $(1.04 \pm 0.10) \times 10^{-4}$ and $(1.52 \pm 0.05) \times 10^{-4}$, see Fig.~\ref{fig:2merged}(e). 

\subsubsection{Measurement: Hong-Ou-Mandel Interference Stations}
At the Charlie station in BNL and Dave station in SBU, active feedback compensates for polarization drifts caused by propagation in the optical fibers (section~\ref{subsubsec:polstab}). Both 1324 nm pulse trains, $\hat{b}_A(t)$ and $\hat{b}_B(t)$, interfere at a 50:50 non-polarizing beamsplitter, $\hat{H}_\text{NPBS} = -(\hbar\theta)(b_{A}b_{B}^{\dagger}+b_{A}^{\dagger}b_{B})$. The beamsplitter outputs, $\hat{c}_1(t)$ and $\hat{c}_2(t)$, are measured at the SNSPD, and the coincidence rate between both interferometer arms is computed.

\subsection{Quantum-Enabling Interface Plane Implementation}

\begin{figure*}
\includegraphics[width=0.7\textwidth]{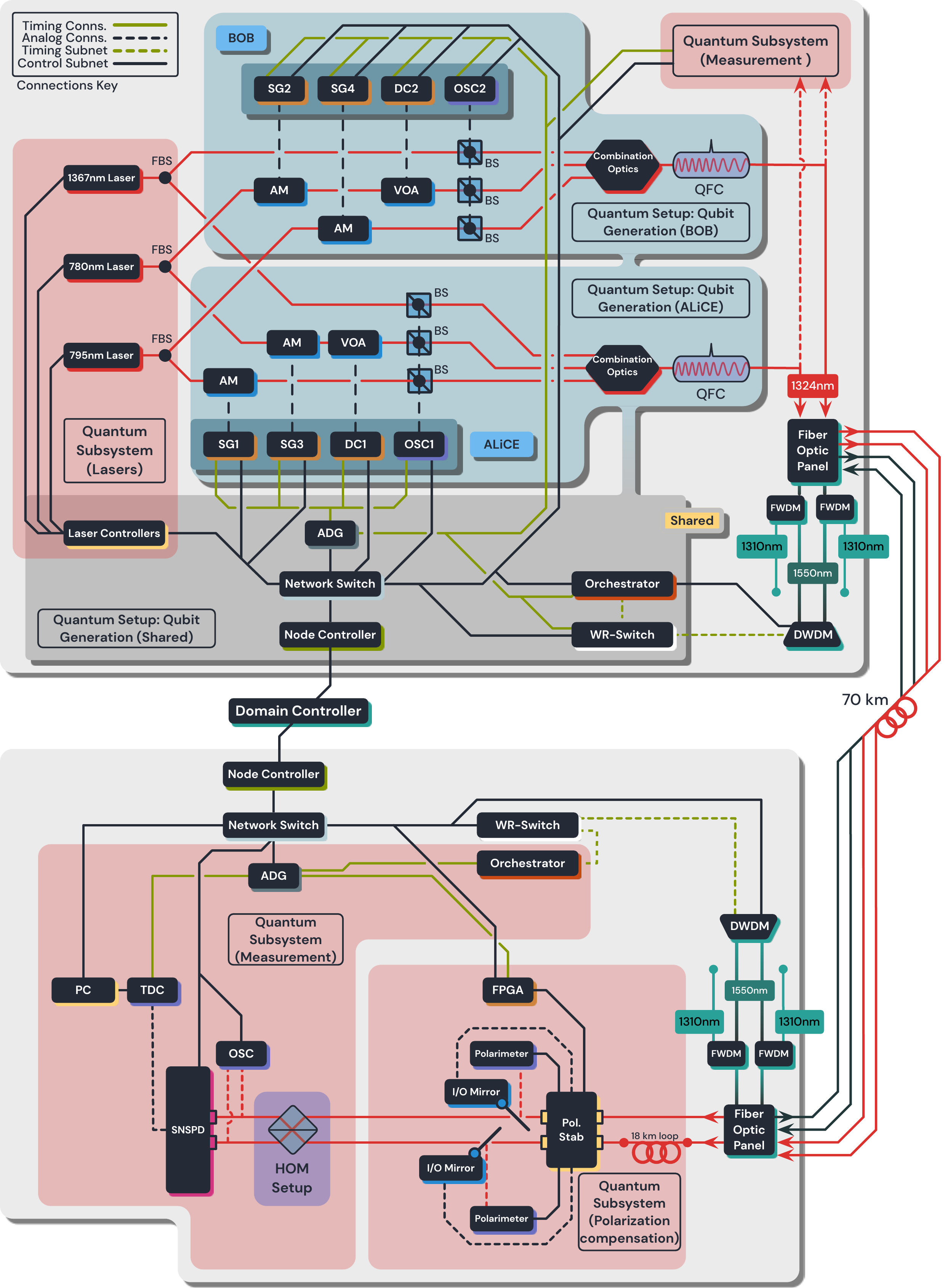}
\caption{\label{fig:HOM_physical}The physical network nodes forming our QEI are shown. At SBU, we have two-qubit generation setups: a laser and measurement subsystem. At BNL, we host a polarization compensation and measurement subsystem. The Alice and Bob qubit generation setups share the control and timing infrastructure (network and timing switches, node controller, orchestrator, multi-channel arbitrary delay generator - gray background) and the laser subsystem (controllers, 795nm, 780nm, 1367nm lasers) distribute laser light to Alice and Bob (red background). Each qubit generation setup hosts a rubidium quantum frequency-conversion system and auxiliary devices (signal generators, delay controller, variable optical attenuator, beam splitters, feedback oscilloscope, combiner) and is connected to a dedicated single-mode long fiber reaching the BNL node (blue background). The SBU local node has a detection subsystem (Dave, right corner) that is identical in functionality to BNL's Charlie node detector subsystem. This configuration adds a 20 km loop in one of the detector input channels. The Charlie node at BNL is configured to host the Hong-Ou-Mandel interference measurement setup. One path from SBU is set up to be longer by approximately 18km, going through BNL's local fiber loop infrastructure for phase randomization between channels (red background). The polarization compensation subsystem compensates for polarization modification in the 70 km. Charlie's detection subsystem hosts an SNSPD, time-to-digital converter, and timing and control infrastructure (red background).}
\end{figure*}

The QEIP provides the functionality necessary for QN protocol execution using deployed fiber infrastructure. Our QEIP consists of five key physical setups and subsystems (see Fig.~\ref{fig:HOM_physical} for relevant physical systems within our QEI): the Alice and Bob qubit generation setups, laser control subsystem, polarization compensation subsystem, and measurement subsystems (see Methods for details). These systems configure and provide the temporal changes in the FWM Hamiltonian required to evaluate the HOM visibility of qubit interference at the end of the quantum channels. In this section, we highlight and discuss the network clocking infrastructure, capable of sub-nanosecond-level synchronization resolution, and the polarization compensation subsystem, providing a scalable approach to making long-haul single-mode fibers compatible with QN operations. 

\subsubsection{\label{subsubsec:timing}Distributed Timing and Synchronization}

The timing infrastructure within our QE interface plane is crucial to ensuring successful network operation, specifically for non-local or interference-based protocols, like HOM, where interference quality quantifies success. To ensure precise time synchronization across distantly located laboratories, we employ the White Rabbit bidirectional time-transfer technology \cite{Lipinski_2011} along 70 km fiber links connecting quantum network nodes. Each node features a WR timing switch delivering 10 MHz analog synchronized clock signals and phase-locked one pulse per second triggering signals.

\begin{figure*}
\includegraphics[width=0.9\textwidth]{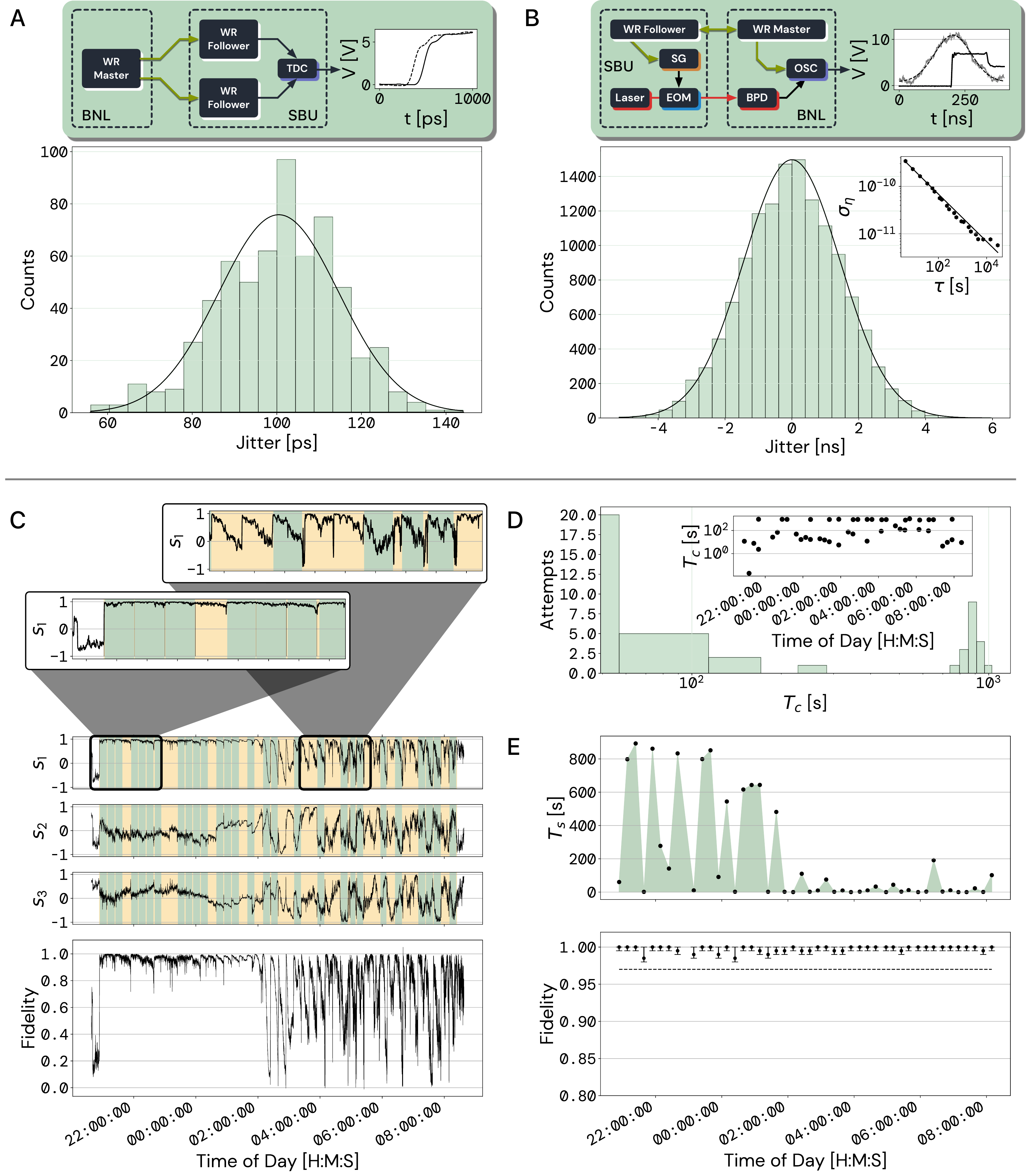}
\caption{\label{fig:4merged}(a) Histogram of jitter events in the 70 km shared clock network, where Charlie (BNL) is the master clock in a tree configuration connecting Alice and Bob (SBU). Bob's time frame shows a jitter of $100\pm 14$ ps with respect to Alice's time frame. Inset: Three-node configuration of the White Rabbit devices and sample delay plot between both SBU clocks. TDC: Time Digital Converter. (b) Histogram of jitter events in the complete SBU-BNL 70 km QN link after 12 hours of integration time. Inset: Allan deviation associated with the long-distance jitter measurement. (c) (Bottom) Distributed polarization states between the SBU and BNL nodes, with compensation executed every 15 minutes over an integration period of 12 hours. The PS is represented with the three normalized Stokes parameters ($S_1, S_2, S_3$) and a Fidelity $F$. The upper plots show a close-up of a stable time segment (around midnight) and a period with more fluctuations (early morning). (d) Distribution of the time needed to stabilize the transmitted PS to desired state $\left|H\right>$. The inset shows no correlation to the time of the day was observed during the 12-hour observation period. (e) (Top) Free-drift time for which the polarization fidelity is preserved over $97\%$ (dotted line in plot below) after the compensation is performed. (Bottom) Polarization fidelity measurements with activated feedback every 15 minutes during a collection time of 12 hours.}
\end{figure*}

To evaluate the performance of our timing system, we measured the phase-locked pulses between the Alice and Bob nodes at SBU, synchronized to the WR source at Charlie in BNL. This setup yielded a network time jitter of $100\pm14$ picoseconds, detailed in Fig.~\ref{fig:4merged}(a). We also characterized the effective network time jitter by including the effect of protocol-relevant QEIP elements. Telecom laser pulses are generated at the SBU node using electro-optical modulators driven by signal generators, in which the local clock is phase-locked to the network timing system WR clock. After propagating over a 70 km fiber, we compare the pulse arrival time at BNL with a similarly clocked signal. Fig.~\ref{fig:4merged}(b) displays a histogram of the timing differences between these pulses over twelve hours. We obtained a Gaussian distribution with an FWHM of $3.4$ nanoseconds, three orders of magnitude smaller than the FWM-generated telecom photons' temporal width.
Furthermore, we estimate the Allan deviation $\sigma_\eta(\tau)$ (see Methods) as a function of varying interrogation times (see the inset in Fig.~\ref{fig:4merged}(b)). We observe the Allan deviation following close a $\tau^{-1/2}$ trend, meaning that over intervals of less than twelve hours, the effective quantum network temporal jitter is stable and predictable. This level of time synchronization allows the control plane to accurately configure the QE interface plane to execute the HOM protocol sequences across our 158 km network and provides a foundation for non-local QN operations.

\subsubsection{\label{subsubsec:polstab}Polarization Compensation}

Maintaining consistent polarization bases between nodes is crucial for ensuring photon indistinguishability; consequently, the polarization compensation procedure is a vital operation executed by our CP and QEIP interface planes. This procedure performs dynamic compensation of polarization drifts across our network to mitigate polarization fluctuations resulting from random birefringence induced by thermal changes, mechanical stress, fiber-core irregularities, and other changing environmental and material inhomogeneities~\cite{Neumann2022, Bostontestbed}. Before conducting HOM measurements using the  QM-QFC generated quantum states, CW 1324nm FWM-generated light is transmitted to characterize and compensate for polarization changes. This compensation is facilitated by a prototype device developed by Qunnect Inc., which employs machine learning to modify the birefringence in compensation fiber loops at the terminals of the 70 km fibers linking SBU and BNL. Feedback from polarimeters informs the compensation procedure. An intermediary FPGA allows the controller to monitor the compensation status and manage the schedule effectively.

We have characterized the complete polarization compensation system over 12 hours as proof of principle, with compensation applied every 15 minutes. We measure polarization fluctuations by obtaining the Stokes vectors of the FWM-generated light during the free-running intermediate times and after executing the compensation procedure. Fig.~\ref{fig:4merged}(c) shows the correlation between the stability of the polarization and the time of day. The compensation is set to correct the initially horizontally polarized FWM-generated light, with initial Stokes parameters $S^0_1=1,$ and $S^0_2=S^0_3=0$. The fidelity of the corrected received signal is calculated as $F=\langle H|\hat{\rho}|H\rangle=(1+S_1)/2$, where $S_1$ is the first Stoke parameter. The fidelity reached after every compensation step is shown in Fig.~\ref{fig:4merged}(e) (Bottom).

Additionally, we have evaluated the network downtime needed to perform the compensation. A histogram showing the statistical distribution of the measured compensation times (every 15 minutes) over 12 hours is shown in Fig.~\ref{fig:4merged}(d), the cluster near $10^3$ seconds represents the failed compensation attempts. Lastly, we have determined the stability times, defined as the times when the fidelity has decreased to $0.97$ after the compensation has been performed. We choose $0.97$ as a lower bound by considering a long-distance HOM service scenario in which two symmetric channels of weak coherent pulses attain a HOM visibility of $\sim\frac{1}{2}f^2\approx 0.47$, where $f$ is the fidelity of the channels with respect to each other \cite{Moschandreou}. These measurements are shown in Fig.~\ref{fig:4merged}(e). We notice that the fidelity routinely remains above our defined threshold for more than 200 seconds during the high-stability periods of the long-distance fiber. Based on these results, we choose to activate the compensation protocol after every three-minute interval of free-drift time for the experiments. This characterization allows us to allows us to attain high HOM visibility during the entire duration of the quantum communication experiments while minimizing the downtime of the network (yellow blocks in Fig.~\ref{fig:4merged}(c)).

\subsection{Control Plane Implementation}

The CP generates commands distributed to all classical and quantum devices across the QEI network through the QEIP and defines the primitive functions necessary to create QN protocols. Our CP addresses the following: (i) using QE-APIs to interface the CP with the QAP and QEIP, (ii) running physics-centric QN primitives associated with the temporal evolution of the quantum variables defining quantum network device performance, and (iii) running physics-centric QN protocols manipulating the Hamiltonian evolution of the quantum devices and measurements.

As an example of this concept, we execute the HOM QN Protocol, a fundamental operation that must be performed to confirm the QN’s readiness to establish an entanglement link. As QN protocols are realized through the interaction of the CP and QEIP, implementing our HOM QN protocol begins by realizing the Southbound QE-APIs. Using the Virtual Instrument Software Architecture (VISA) protocol and custom device communication standards, we developed QE-APIs consisting of Python libraries to address and control network devices homogeneously, depending on functionality. These QE-APIs enabled the development of the eight-step software procedure used by the control plane to orchestrate the QEI's operations for running the HOM protocol. These primitives/operations run on the QEI to extract photo-counting measurements at the end node, from which we can verify high visibility HOM interference. These QN primitives include the modification of parameters of the frequency conversion FWM Hamiltonian, $\pi_\text{Alice/Bob}=\{\Omega_\text{I}(t),\ \Omega_\text{II}(t),\ \langle \hat{a}_{A,B}^{\dagger}(t)\hat{a}_{A,B}(t)\rangle \}$ (see Eq.~(\ref{eq:dipPulsed})).

\subsubsection{Execution of the HOM Protocol}

\begin{figure*}
\includegraphics[width=1.0\textwidth]{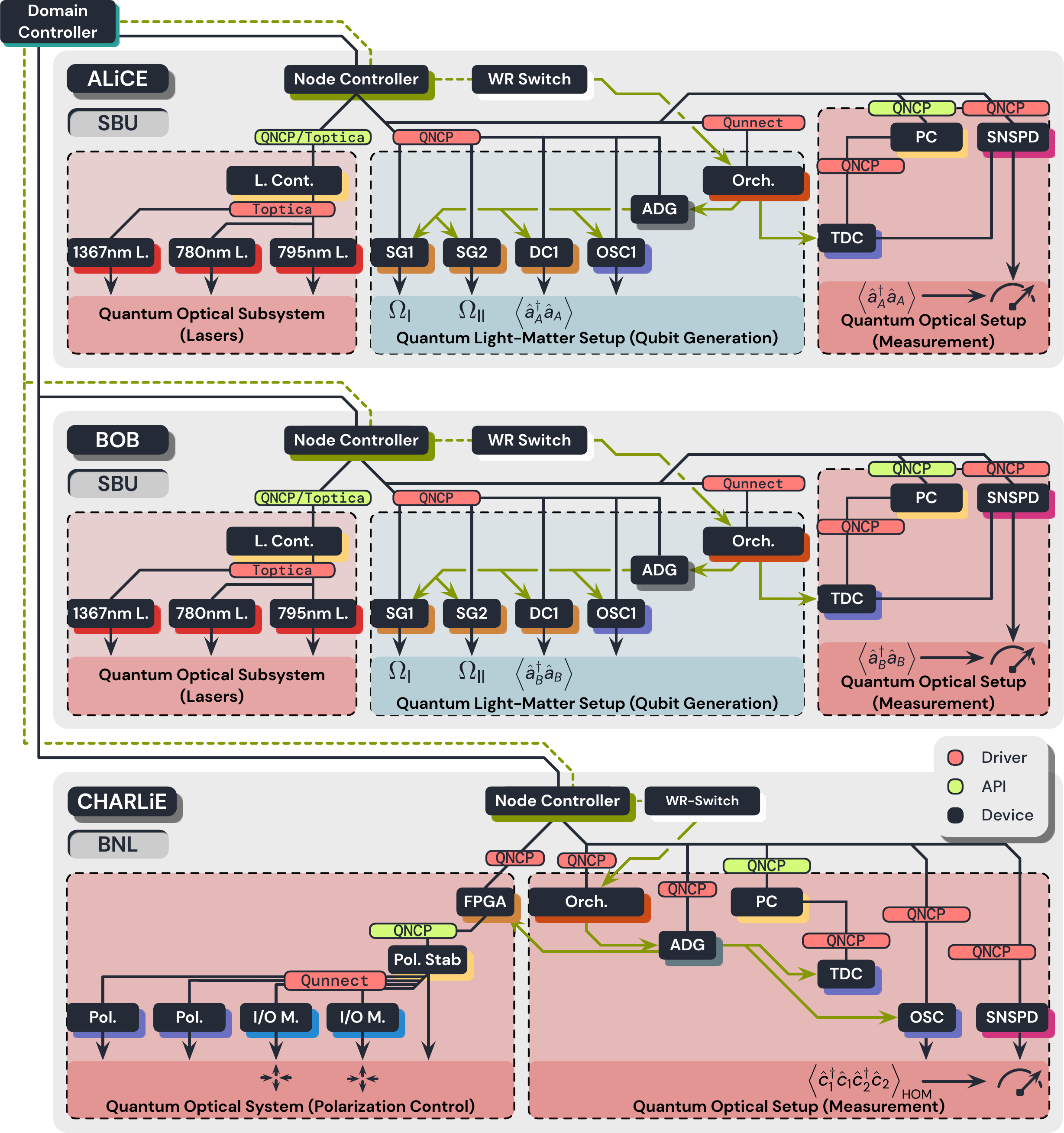}
\caption{\label{fig:HOMqe}The implementation of the QE interface plane: the BNL/SBU QEI testbed has two identical co-located nodes at SBU and a third node at BNL. The figure shows the devices of each node (here depicted individually), color-coded by functionality, their connectivity, and their grouping in quantum optical system setups (blue background) and quantum optical subsystems (red background).}
\end{figure*}

The execution of QN primitives and protocols highlights the coordination between the QDP, QEIP, and CP. This coordination, in the context of the full-network Hong-Ou-Mandel (HOM) visibility-check protocol, is presented in Fig.~\ref{fig:HOMqe}. To perform the HOM indistinguishability protocol, the CP must coordinate and optimize the function and interplay of the QN primitives, which is done using physics-relevant subsystems within the network via the QE-protocols. Fig.~\ref{fig:HOMqe} presents a streamlined view of our network hierarchy between three nodes and describes how the controller adjusts the active network between QE protocols, thereby facilitating the execution of the QN primitives forming our QN protocol.

\begin{figure*}
\includegraphics[width=\textwidth]{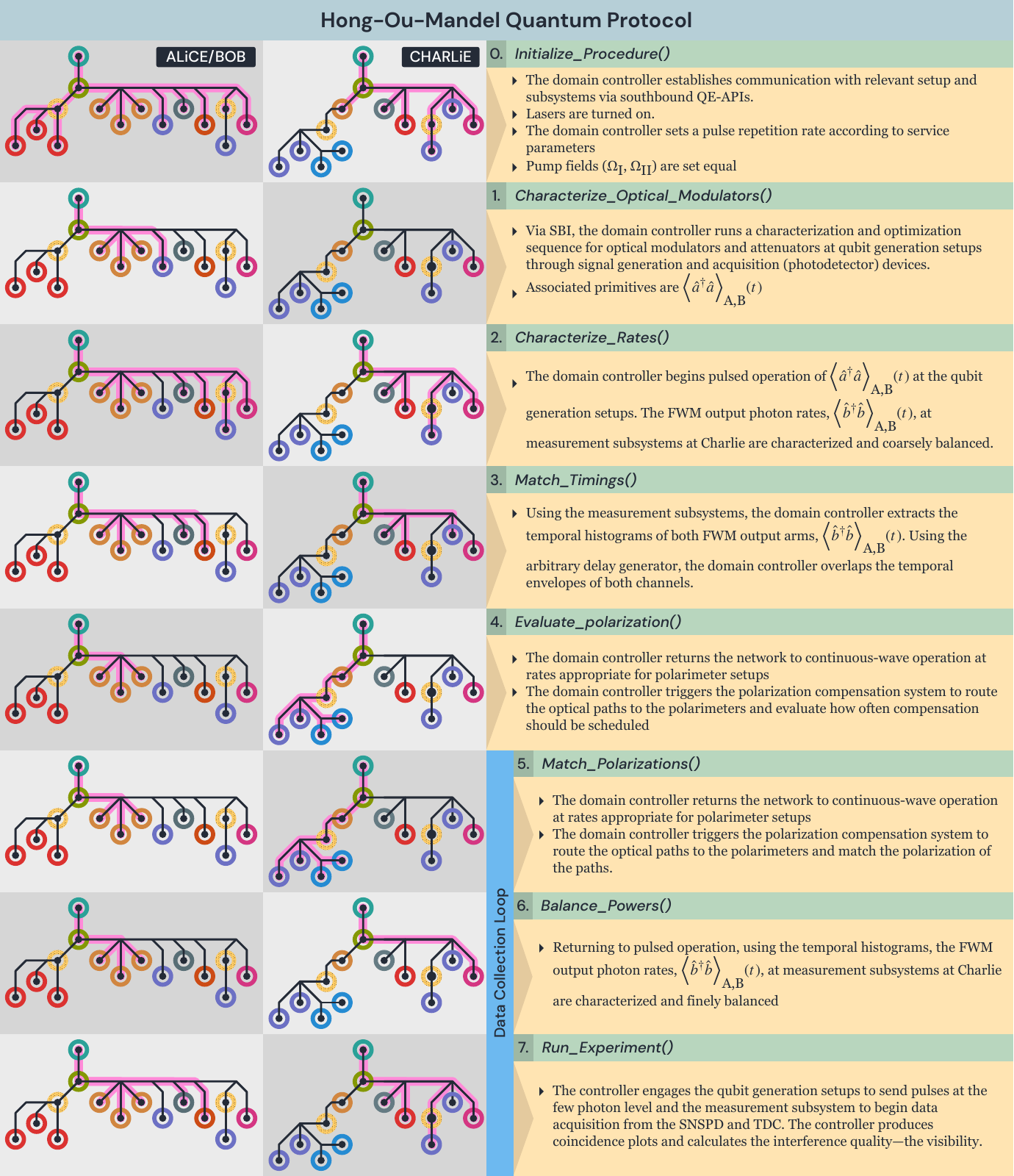}
\caption{ \label{fig:HOMprotocol}The HOM protocol and highlighted network primitives executed by active QEIP components.}
\end{figure*}

The (HOM) visibility-check protocol begins by ensuring that our central controller can communicate with network elements corresponding to the Hamiltonian parameters in the FWM systems: the Alice and Bob qubit generation (controlling the temporal evolution of the quantum variables $\langle \hat{a}_{in}^{\dagger}(t)\hat{a}_{in}(t)\rangle$ and  $\langle \hat{b}_{in}^{\dagger}(t)\hat{b}_{in}(t)\rangle$), the control-fields control (controlling the time evolution of the variables $\Omega_I(t)$ and $\Omega_{II}(t)$ in the FWM Hamiltonian for each QM-QFC), the polarization stabilization subsystem (guaranteeing the high fidelity preservation of the quantum variables $b_{Alice}$ and $b_{Bob}$ after long-distance propagation), and the HOM measurement subsystem (measuring the overlap between the quantum variables $\langle \hat{a}_{HOM}(t) | \hat{b}_{HOM}(t)\rangle$) (see Fig.~\ref{fig:HOMqe}). The interplay between the evolution of the QN primitives and the execution of the HOM QN protocol is detailed in Fig.~\ref{fig:HOMprotocol}. 

\subsection{Quantum Application Plane Implementation}
A QN service is executed by physically running QN protocols, using the interaction of all the network planes via the QAP. Users or higher-order network processes initiate quantum services through an NBI, communicating requests to the CP. Based on the QN service requirements, the CP then issues a sequence of commands to the QEIP via an SB QE-API, enabling the physical realization of the QN service at the QDP. 

\subsubsection{High-visibility Hong-Ou-Mandel Quantum Network Service}

A first example of a robust QN service enabled by our QEI is the verification of high-visibility HOM interference over long distances using light-matter quantum interfaces. This capability is a ubiquitous and essential requisite for implementing more advanced QN protocols such as Cabrillo entanglement generation, Duan-Lukin-Cirac-Zoller entanglement creation, memory-assisted entanglement swapping, and quantum-repeater-assisted teleportation. Successfully executing high-visibility HOM interference at distant network nodes serves as a crucial measure of the network’s ability to distribute entanglement and perform teleportation, as demonstrated by the Peres Horodecki criterion~\cite{Peres1996} and the CHSH bound. These criteria link HOM visibility directly to the quality of entanglement generation and delivery~\cite{Clauser1969}.

To benchmark the quality of a QN service, it is necessary to establish a framework that relates the network flow to the performance of a quantum measurement operation. For the particular case of the long-distance high-visibility HOM QN service, we compute a second-order correlation between the outputs of the beamsplitter at Charlie or Dave and compare it to an analytic model to quantify the visibility of the interference effect (See derivation in Supplemental Material). Taking our pulse train and frequency lineshapes to be Gaussian, we can evaluate our interference visibility using the following:

\begin{widetext}
\begin{equation}
\begin{aligned}
\frac{dN^{(2)}}{d \, \Delta t}  \;  &\propto  \;
    \left[\sum_n G(\Delta t; n\Delta T,\sqrt{2}\sigma_t)\right] - \mathcal{V} G\left(\Delta t; 0, \frac{\sqrt{2}\sigma_t}{\sqrt{1+ (4 \pi \sigma_\nu \sigma_t)^2}}\right) \cos\left(\delta \omega \Delta t\right)
    \label{eq:dipPulsed}  
\end{aligned}
\end{equation}
\end{widetext}

Where $G(t;t_0,\sigma) = \exp(-(t-t_0)^2/2\sigma^2)$, $\Delta t$ is the temporal difference between events being correlated, $\Delta T$ is the pulse repetition period, $\sigma_{t}$ and $\sigma_{t}$ characterize the temporal pulse and light frequency of the incoming pulses, $\delta\omega$ is the central frequency difference between inputs and $\mathcal{V}$ is the interference visibility.

We can see from Equation~(\ref{eq:dipPulsed}) the essential feature that HOM interference will deplete the pair rate within the ``central'' peak centered at $\Delta t =0$. The pulse modulation determines the linewidth when the pulses are much shorter than the CW coherence time. The central peak in the pairs' distribution is scaled down by a factor of $(1-\mathcal{V})$. For the intermediate regime where the pulse width and CW coherence time are comparable, the central lobe of the distribution of the pairs can take on a two-peaked shape, as seen in the data below. By fitting the HOM experimental data and obtaining the figure of merit $\mathcal{V}$, the HOM QN service is completed by verifying the high-visibility HOM interference over long distances 

\section{\label{sec:experiments}Long-Distance High-Visibilty HOM QN Service}

\begin{figure*}
\includegraphics[width=\textwidth]{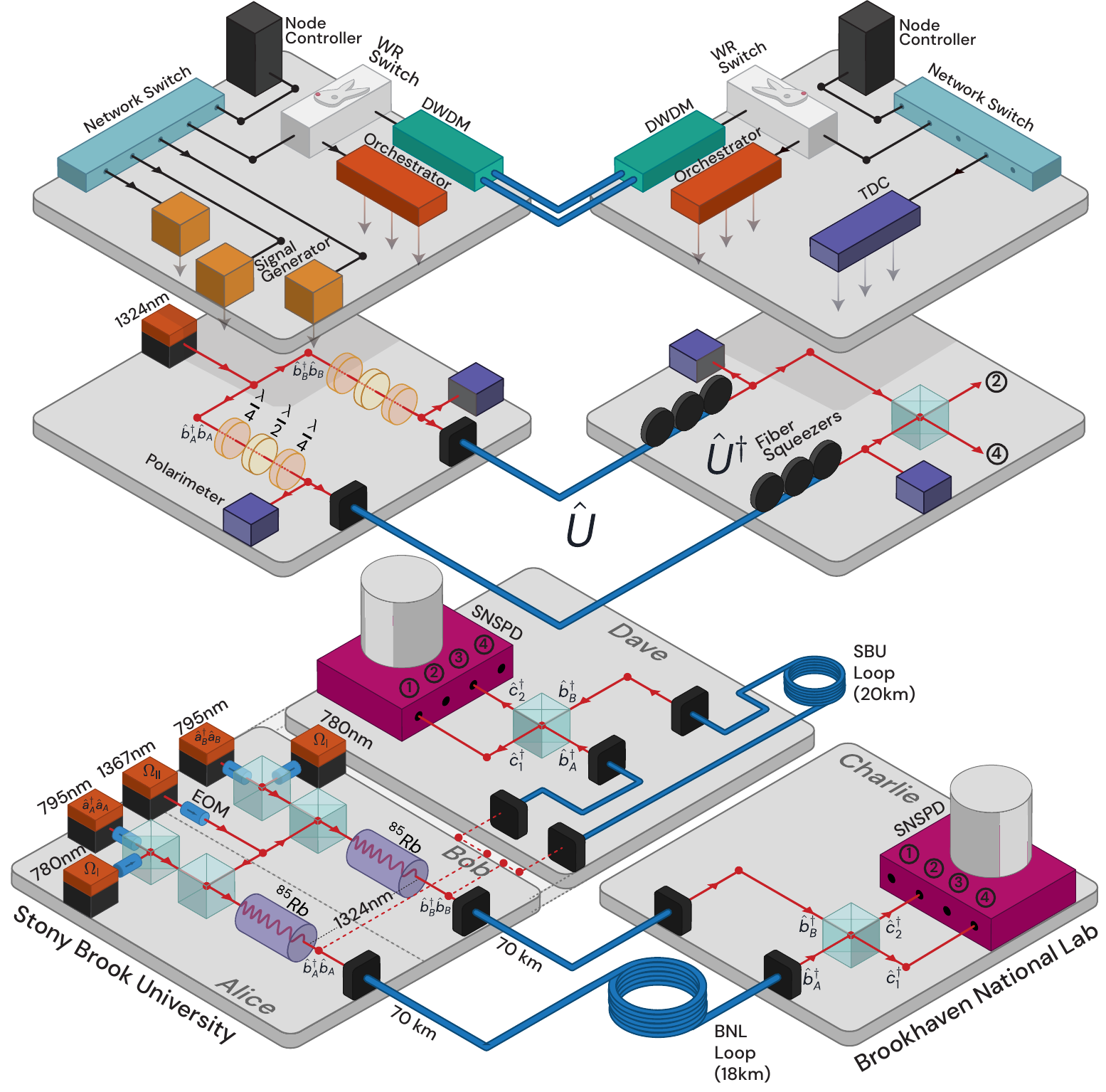} 
\caption{\label{fig:QNsetup2}Overview of the four-node QN used in the long-distance high-visibility HOM QN service. Two quantum memories and frequency conversion setups (Alice and Bob) are co-located at SBU and are connected independently to the network. The interference setups and telecom-compatible single photon nano-wire detectors (Charlie and Dave stations) are located at BNL and SBU, respectively. Optical fibers transport quantum information (bottom layer) and quantum-enabling information (middle layer). Other fibers transport classical timing triggers, network status, and sequencing information (top layer). The layering shown represents the coexistence of the three kinds of information needed for the function of our quantum network: classical, quantum-enabling, and quantum.}
\end{figure*}

\begin{figure*}
\includegraphics[width=0.9\textwidth]{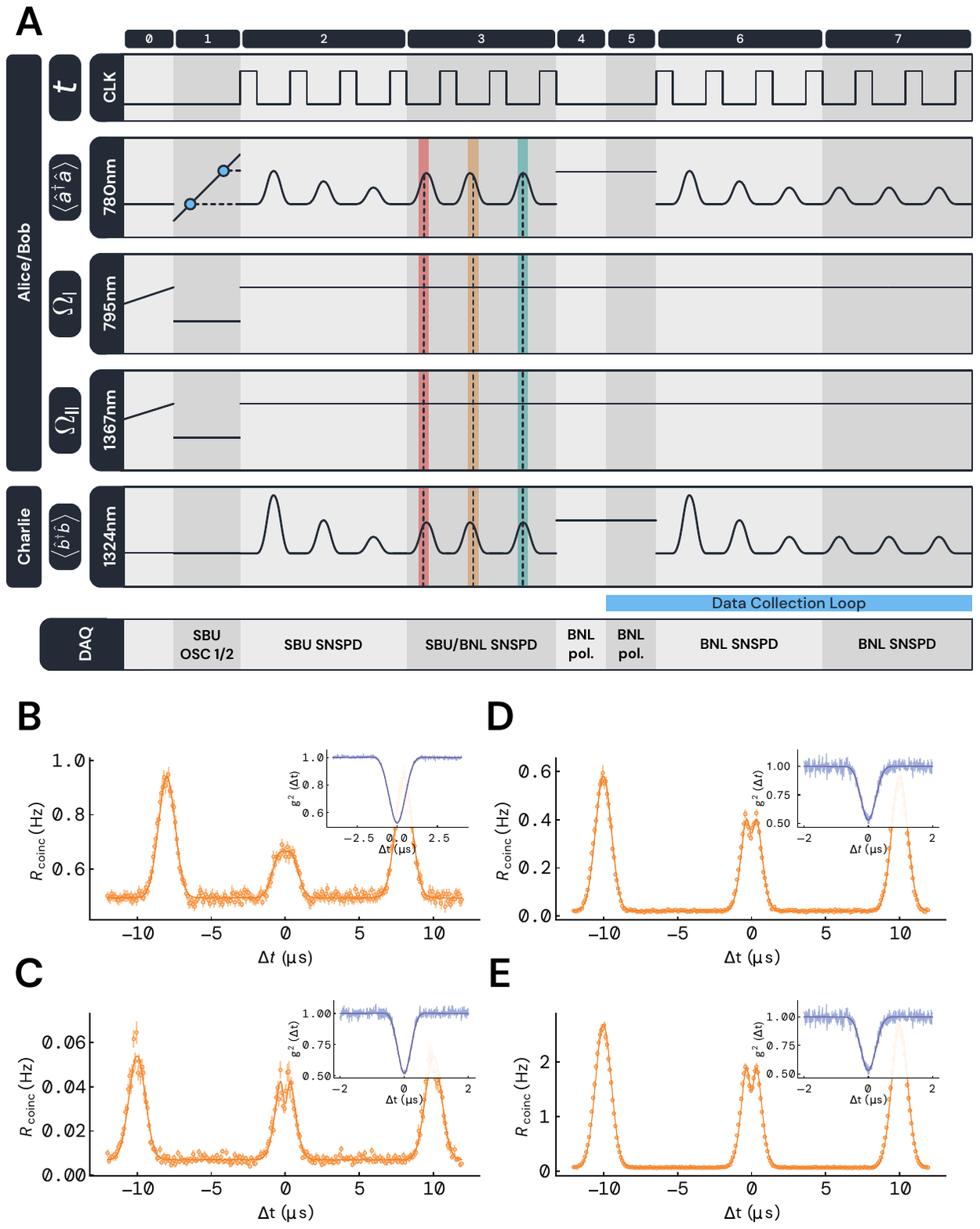}
\caption{\label{fig:merged8}(a) Orchestrated sequence for QM QFC generation and long-distance HOM interference. We show the time dynamics of each node and their devices within the network, as depicted in the Methods section. (b) The 1324 FWM HOM pulsed quantum interference experiment operated for the 20 km QEI configuration at the SBU site. Inset: Continuous-wave HOM. (c)-(e) The 1324 FWM HOM pulsed quantum interference experiments operated for the 158 km QEI configuration at the SBU and BNL sites. The measurements are done for different mean photon numbers at the exit of the frequency converters, see TABLE~\ref{table_pulseHOM} for detail data. The errors are calculated as the standard deviation of twenty mean values. Each data set corresponds to collecting data for 3 minutes (18 million pulses sent).}
\end{figure*}

We now execute the first HOM QN service in our layered 158 km QEI(see Fig.~\ref{fig:QNsetup2} for details). For characterization purposes, we start by using an experimental configuration within the SBU campus, with an HOM measurement station at the Dave node. This configuration uses a fiber loop communicating the main Physics building to the CEWIT campus over a distance of 20 km (SBU Loop in Fig.~\ref{fig:QNsetup2}), to perform the HOM visibility-check QN service. These experiments are driven using the QEI architecture to create independent quantum state sequences in each QM-QFC. The sequencing is controlled using the network primitives defined above (see Fig.~\ref{fig:HOMprotocol}), in the following manner (see Fig.~\ref{fig:merged8}(a) for a complete sequence):

\paragraph{} characterization of the 780 nm qubits modulation using the CP and QEIP to execute the \textit{Initialize\_Procedure()} and \textit{Characterize\_Optical\_Modulators()} QE protocols.
        
\paragraph{} application of QFC pumps to produce 1324 nm photon streams and verification of QM QFCs output mean photon numbers using the QDP, CP and QEIP to execute the \textit{Charaterize\_Rates()} QE protocol, characterizing the QN primitive associated to $\left<\hat{a}^\dagger\hat{a}\right>$. This creates a pulse repetition rate of 125 kHz, and a temporal envelope $\sigma_{t}=0.4\mu$s.
        
\paragraph{} preparation of long-distance HOM data collection over thousands of production cycles using the  QDP, CP, QEIP and QDP to execute \textit{Match\_Timings()} and \textit{Balance\_Powers()} QE protocols. This produces a mean photon rate was $\sim$ 800 kHz for both Alice and Bob conversion channels at the HOM measurement beam splitter at the Dave SBU station, preparing the primitive associated to $\left<\hat{b}^\dagger\hat{b}\right>$.
        
\paragraph{} verification of polarization preservation across the long-distance network using the CP an QEIP to run the \textit{Evaluate\_polarization()} and \textit{Match\_Polarizations()} QE protocols.
        
\paragraph{} quasi-real time long-distance HOM coincidence analysis using the  QDP, CP, QEIP and QDP to execute \textit{Run\_Experiment()} QE protocol. This creates a mean number of photons at the input ports of the HOM BS is $\rm \langle n \rangle\approx$0.04/pulse for both Alice and Bob. 

Fig.~\ref{fig:merged8}(b) shows the result of the HOM QN service at the Dave node. The inset shows the result for a CW input and the main figure shows the pulsed results, exhibiting the main features predicted by our theoretical model. The figure also shows the fits of the experimental data to Eq.~(\ref{eq:dipPulsed}), from which a visibility of 47\% can be determined, showing the robustness of the QN service over a distance of 20 km.

\subsection{\label{sec:longdistanceHOM}Stack-Controlled 158 km High-Visibility HOM Service}

We now demonstrate the degree of indistinguishability of the telecom polarization qubits transduced in two independent QM-QFCs using HOM interference experiments, with one arm of the interferometer being 70km and the second being 88km. Distance-wise, these experiments are already competitive with realizations shown recently \cite{Bock2018, Leent2020,Yu2020}.

Fig.~\ref{fig:merged8}(c) shows the result of the HOM QN service over 158 km for a pulsed input with a mean number of photons at the output of each QM-QFC of $\rm \langle n \rangle_{Alice}=55.5\pm2.8$ and $\rm \langle n \rangle_{Bob}=9.5\pm0.5$ defined within one Gaussian pulse temporal envelope FWHM of $\sim 0.9\,\mu s$. The total losses (QM-QFC output to SNSPD input) through both the long-distance setups are measured to be $\rm \sim 28.3 dB$ for one link and $\rm \sim 36.0 dB$ for the longer arm. After long-distance propagation, we obtain an average photon number per pulse of $\rm \langle n \rangle\approx0.014$ for both Alice and Bob conversion channels at the HOM measurement beam splitter in BNL.  The HOM coincidence rate is measured versus the arrival time of the photons in the two detectors. The coincidences within a temporal region of interest (ROI) are post-selected with a width of 93.4 ns (24 $\mu$s divided into 257 bins). We observe the desired modulation in the coincidence rate, exhibiting a minimum for the initial identical polarization and reaching a maximum corresponding to uncorrelated photons beyond the coherence time of the FWM process. The figure also shows the fits of the experimental data to Eq.~(\ref{eq:dipPulsed}). An interference HOM visibility is measured to be $\mathcal{V}=(47\pm4)\%$and an FWM spectral width of $\sigma_\omega=(2\pi)\times(609\pm91)$ kHz. The errors are calculated as the 68\% confidence intervals of the fitting curves. Before collecting pulsed HOM data, we verified the HOM setup alignment as a CW HOM service. Fig.~\ref{fig:merged8}(c)-(e) (inset) shows the result of the HOM QN service over 158 km for a CW input with mean photon rates of $\sim330$ kHz for both Alice and Bob conversion channel at the HOM measurement beam splitter in BNL. The figure also shows the fits of the experimental data, from which a HOM visibility of $48\%$ and a bandwidth of $\sigma_\omega=(2\pi)\times453$ kHz can be determined. 

Fig.~\ref{fig:merged8}(d)-(e) are the results the HOM QN service over 158 km for different input photon numbers per pulse at the output of the QM QFCs. The pulse repetition rate was 100 kHz for all three cases, and statistics analysis was carried out with one hour of integration time for each case. Polarization compensation was performed every three minutes. The main data obtained from these measurements are summarized in Table~\ref{table_pulseHOM}. From the data fits, we obtain the characterization of the quality of the HOM QN service. We obtain photon spectrum FWHM widths of $1.4\pm 0.2$ MHz, $1.07\pm0.06$ MHz, and $1.21\pm0.06$ MHz, with corresponding photon FWHM pulse widths of $0.92\pm0.01 \,\mu$s, $0.889\pm 0.004 \,\mu$s, and $0.880\pm0.002 \,\mu$s, probing the repeatability of the stack-driven HOM QN service. Most importantly, we have measured HOM visibilities of $(47\pm 4)$\%, $(43\pm1)$\%, and $(47\pm1$)\%, showing that the good visibility of the service is preserved as we approach a true single photon level at the output of the QM-QFCs. 

\begin{table}[b]
\caption{\label{table_pulseHOM}
Data from the three pulsed HOM services are displayed in Fig.~8. $\langle n \rangle_{Alice, SBU}$ and $\langle n \rangle_{Bob, SBU}$ are the estimated number of photons per pulse leaving the SBU node. $\langle n \rangle_{BS}$ is the number of photons per pulse of each quantum channel arriving at BNL HOM measurement NPBS input ports. $\mathcal{V}$ is the visibility from the fitting from Eq. (5). $R_{coinc}$ is the coincidence rate at $\Delta t = 0$ obtained from the fit. Errors are calculated as the standard deviation of the means of twenty data sets.}
\begin{ruledtabular}
\begin{tabular}{cccc}
\textrm{}&
\textrm{HOM1\footnote{Fig.~\ref{fig:merged8}(c)}}&
\textrm{HOM2\footnote{Fig.~\ref{fig:merged8}(d)}}&
\textrm{HOM3\footnote{Fig.~\ref{fig:merged8}(e)}}\\
\colrule
$\langle n \rangle_{Alice, SBU}$  &  55\,$\pm$\,3  &  183\,$\pm$\,9  &  387\,$\pm$\,24 \\[0.5ex]
$\langle n \rangle_{Bob, SBU}$ & 9.5\,$\pm$\,0.5 & 31\,$\pm$\,2 & 66\,$\pm$\,4 \\[0.5ex]
$\langle n \rangle_{BS}(\times 10^{-2})$ & 1.40\,$\pm$\,0.07 & 4.6\,$\pm$\,0.2 & 9.8\,$\pm$\,0.6 \\[0.5ex]
$\mathcal{V}(\%)$ & 47\,$\pm$\,4 & 43\,$\pm$\,1 & 47\,$\pm$\,1 \\[0.5ex]
$R_{coinc}$(Hz)  &\ 0.033 \ &  \ 0.34 \ & \ 1.48 \\[0.5ex]
\end{tabular}
\end{ruledtabular}
\end{table}

\section{\label{sec:conclusions}Conclusion}

From these experimental measurements, we can evaluate the performance of our QEI if we were to apply the original Cabrillo scheme to generate quantum memory entanglement over long distances \cite{Cabrillo1999}. The first figure of merit is the long-distance HOM visibility measured at the Charlie node for a single-photon-level output at the QM-QFC sites. We estimated this from the data presented in Table I to be approximately $47\%$. Assuming further reductions in the detector noise and that single-photons have the same fidelities as coherent states \cite{Michelberger2015}, we would expect a single-photon HOM visibility of 94\%, far surpassing the Peres Horodecki criterion of 33\% limit for entanglement generation \cite{Peres1996} and also beyond the 71\% CHSH bound to use matter-matter entanglement to perform Bell inequality violation experiments \cite{Clauser1969}. The second figure of merit is the two-photon coincidence rate for single-photon-level inputs, which, from the data presented in Table I, can be estimated to have a lower bound of approx. $4\times10^{-4}s^{-1}$ for the long-distance experiments. This estimation is competitive with the two-photon rate measured in state-of-the-art long-distance matter-matter entanglement experiments \cite{Yu2020, Neumann2022, vanLeent2022, EnsemblesEntangled2022}. 

%\section*{Outlook}\label{sec:outlook}
These results make us believe that our QN architecture paradigm and current QEI infrastructure can be extended to perform quantum memory entanglement experiments over unprecedented distances in deployed fibers. These extensions will include the QAP running the QN service of entanglement generation among the quantum memories, requiring two significant additions. First, at the QDP level, we will extend our QFC scheme to generate entanglement with photons directly at 1324 nm and the collective state of the QM. This can be done by correlating the polarization state of the created 1324 photon to that of the resultant intermediate magnetic atomic sublevels \cite{uphoff2016integrated}. Second, after the photon's propagation in the long-distance links at the CP level, we will perform a Bell-state projection QN service, regulating the creation of entanglement among the magnetic substates of the remote QMs. This measurement will require further additions to our QEIP, including having QE-APIs running unperturbed over several hundreds of hours and phase stabilization across the whole quantum network testbed, which can be achieved by monitoring the long-distance HOM interference service shown here \cite{Fang2022}.

An additional improvement to our QDP will be to have long QM coherence times on the order of milliseconds \cite{dideriksen2021room}, allowing for the long-distance transmission of the telecom photons and the Bell state result transmission across the quantum-enabling network. With these conditions, it will be possible for the QAP to verify the entanglement created between the QMs by retrieving their state into an entangled photonic state and reconstructing its density matrix. 

In conclusion, we have presented the design, deployment, and implementation of a first instance of a QEI network connecting QM-QFC atomic ensembles over 158 km of deployed fiber. Using our novel QN design paradigm, we have demonstrated a stack-driven ubiquitous QN service, delivering long-distance, robust HOM interference with high visibility.

In the near future, we envision applying this QN paradigm design to demonstrate further scalable long-distance QN services in our current three-node QN infrastructure, including the transmission of polarization entanglement created by high repetition sources of entangled photons and the storage of telecom polarization entanglement using remotely located quantum memories capable of heralding the storage of entanglement using non-demolition measurements and quantum state tomography. These advanced experiments will lay the foundation for a stack-driven, long-distance quantum repeater implementation.

\begin{acknowledgments}
The authors thank Dr. Olli Sera, Dr. Steven Sagona, and Dr. Mehdi Namazi for their contributions at the early stages of the manuscript. We gratefully acknowledge Mael Flament and Gabriel Bello of Qunnect Inc. for their comprehensive technical support on the polarization compensation device and timing distribution. We also thank Prof. Connor Kupchak and Dr. Joanna Zajac for fruitful discussions and comments on the manuscript. Additional thanks to Anthony del Valle and Samuel Woronick for their technical assistance in the later stages of the experiment. This work was supported by the DOE ASCR grant KJ0402030: ``Inter-campus network enabled by atomic quantum repeater nodes'' and the DOE CESER grant M620000044: "A Prototype for Ultimate Secure Transmission and Analysis of Smart Grid Data on the Wire". D.D., L. C. V., D. K., and E.F. acknowledge support from the Center for Distributed Quantum Processing at Stony Brook University. D. K. and J. M.-R. acknowledge support from the Instrumentation Department at Brookhaven National Laboratory. 
\end{acknowledgments}
\vspace{\baselineskip}
%M. F. and E. F. have shares of Qunnect Inc., a quantum technology company based in NY. All other authors declare no competing interests. 

\bibliography{scibib}% Produces the bibliography via BibTeX.

\newpage

% Reset counters for the second file
\setcounter{section}{0}
\setcounter{equation}{0}
\setcounter{page}{1}
\renewcommand{\theequation}{S\arabic{equation}}

\begin{center}
    \Large\bfseries Supplementary Materials: A Long-distance Quantum-capable Internet Testbed\\
\end{center}

\section{Methods}

\subsection{Frequency locking of 1367 nm laser:} See Fig.1 in main text. We use the $\rm 5S_{1/2}\rightarrow5P_{3/2}$ transition ($\ket{1}\rightarrow\ket{3}$) at $\rm 780nm$ and the $\rm 5P_{3/2}\rightarrow6S_{1/2}$ transition ($\ket{3}\rightarrow\ket{4}$) at $\rm 1367nm$. The $\rm 780nm$ optically pumps the $\rm 5P_{3/2}$ level, and it is locked to the $\rm 5S_{1/2}F=3\rightarrow5P_{3/2}F'=4$ transition via saturated absorption spectroscopy (SAS). For OODR, the $\rm 1367nm$ laser is counter-propagating to the 780 nm laser beam in a $\rm 7cm$ rubidium cell maintained at $\rm 60^{\circ}C$. An InGaAs balanced amplified photodetector obtains the OODR spectrum when scanning the $\rm 1367nm$ laser while keeping the 780 nm laser locked. An error signal is generated by modulating the $\rm 1367nm$ laser at $\rm 100kHz$, and a PI controller locks the laser according to the error signal.\\

\subsection{Quantum-Enabling Application Programming Interfaces:}

Our Quantum-Enabling (QE) Application Programming Interfaces (APIs) define how the control plane communicates with our QE interface and application planes. Southbound QE-APIs consist of drivers that handle communication between the controller and all devices and sub-system controllers on the QE interface plane. These drivers are python-written and based on the Virtual Instrument Software Architecture (VISA) protocol and custom device communication standards. Depending on functionality, these libraries follow a facade design pattern to control network devices homogeneously. Our eight QE-protocol step procedure is built from these drivers.

Our northbound QE-APIs are Python commands written on top of our eight-protocol procedure. They can extract measurement and status information, including mean photon number, temporal photon shapes, and HOM visibility, along all procedure steps. We developed a dashboard/GUI to use the northbound QE-APIs to present relevant information as the QN protocol is executed.

\subsection{Controllable Quantum Network Nodes:} We identify four controllable nodes in our QEI network. Alice and Bob are light-matter interface nodes located on SBU campus. The two nodes share the same laser subsystem including lasers at 795nm, 780nm, and 1367nm wavelengths and associated locking systems. The shared lasers are split and sent to independent QM-QFC setups in Alice and Bob. Each QM-QFC setups have its independent acousto-optical modulators, electro-optic modulators and optical attenuators to carry out primitive operations.

The SBU nodes share a network/timing switch, controller, sequence orchestrator, and multi-channel delay generator. The QM QFC nodes also share a laser subsystem, including lasers at 795nm, 780nm, and 1367nm wavelengths, controlled by a dedicated computer. The QFC qubit creation procedure is executed via acousto-optical devices and variable optical attenuator (VOA) devices that modulate each laser. We use feedback to optimize the signal-to-background of the amplitude modulation. The modulated laser signals are combined in an optical setup and then sent to the QM QFC systems. Their outputs are connected to dedicated fibers that carry the quantum signals to the BNL node. 

 At the BNL (SBU) measurement nodes the quantum states arrive through the long fibers and are directed to an HOM detection subsystem equipped with polarization stabilization subsystems. The control of all the BNL node components is done with a node controller and a network switch, while the synchronization is done with a timing switch, a sequence orchestrator, and a delay generator. One optical path from a QM QFC system is set up to be longer by approximately 18km (20km) of fiber, going through BNL's (SBU) local fiber loop infrastructure for phase randomization. The polarization stabilization system uses computer-controlled mirrors to direct macroscopic QFC states to polarimeters for measurements during the stabilization phase. Once the appropriate compensation has been calculated, the mirrors are automatically removed from the path, allowing the photons to go to the HOM detection setup. The HOM setup utilizes two SNSPD channels connected to a Time-to-Digital Converter (TDC)  supervised by a dedicated PC and driven by a signal generator. The detector is also connected to the classical network, allowing real-time status monitoring and data acquisition.\\

\subsection{Allan Deviation of QN clock jitter:} We define the normalized jitter as $\eta_i=1+f_R\times t_i$, where $f_R=10$ kHz is the pulsing repetition rate and $t_i$'s are the jitter measurements. We define the Allan variance as
\begin{equation}
\sigma^2_\eta(\tau)=\frac{1}{2(M-1)}\,\sum_{k=1}^{M-1}\left(y_{k+1}-y_k\right)^2,   
\end{equation}
where $y_k=(1/L)\sum_{i=1}^L \eta_{(k-1)L+i}$, $k=\{1,2,...,M\}$, $L$ is the number of data points in each segment of interrogation time $\tau$, $L=\tau f_R$, $T=M \tau$, and $N=M L$. $T$ is 12 hours and $N$ is the total number of collected data points.\\

\section{Theory of quantum frequency conversion in Rubidium}
In our quantum frequency conversion system, we are using hot rubidium vapor in a glass cell. We model the system with an atomic ensemble in a cavity and apply the input-output theory(\textit{54}, \textit{41}). In the end, we go to the bad cavity limit. The Hamiltonian of the system is described as $\hat{H} = \hat{H_0} +\hat{H_L}$, with the unperturbed Hamiltonian defined as

\begin{equation}
\begin{aligned}
\hat{H_0} &=\sum^N_{i=1}\left[\hbar\omega_2\sigma^i_{22}+\hbar\omega_3\sigma^i_{33}+\hbar\omega_4\sigma^i_{44}\right]+\hbar\omega_{a}a^{\dagger}a+\hbar\omega_{b}b^{\dagger}b,
\end{aligned}
\end{equation}
where $i$ represents different atoms in the ensemble. We set the ground state of the atom as the zero energy point. Here $\sigma_{ii} = \ket{i}\bra{i}$, $a$ and $b$ represent 780nm mode and 1324nm mode inside the cavity. The interaction term under dipole approximation takes the form 
\begin{equation}
\begin{aligned}
\hat{H_L} &= \sum_i^N-\hat{\textbf{d}}^i\cdot\textbf{E}=\sum_i^N-q\hat{\textbf{r}}^i\cdot\textbf{E}\\
\end{aligned}
\end{equation}
In the current four-level Hilbert space (See Fig. 4 in the main text),
\begin{equation}
\hat{\textbf{d}}^i = \textbf{d}_{12}\sigma^i_{12}+\textbf{d}_{13}\sigma^i_{13}+\textbf{d}_{24}\sigma^i_{24}+\textbf{d}_{34}\sigma^i_{34}+\text{h.c.}
\end{equation}
where we choose the phase of atomic states $\ket{i}$ so that the dipole matrix element is real. We treat the pump fields as classical fields, and we quantize the two weak fields such that
\begin{equation}
\begin{aligned}
\textbf{E} &= \textbf{E}_{I}+\textbf{E}_{II}+\textbf{E}_s+\textbf{E}_t,\;\textrm{with}\\
\textbf{E}_{I} &= \vec{\epsilon}_I\left(\frac{\mathscr{E}_I}{2}\right)e^{-i\omega^L_It+i\vec{k_{I}}\cdot \vec{r}}+\text{c.c.}\\
\textbf{E}_{II} &= \vec{\epsilon}_{II}\left(\frac{\mathscr{E}_{II}}{2}\right)e^{-i\omega^L_{II}t+i\vec{k_{II}}\cdot \vec{r}}+\text{c.c.}\\
\textbf{E}_{s} &= \vec{\epsilon}_{s}\mathscr{E}_{s}\hat{a}e^{i\vec{k_{s}}\cdot\vec{r}}+\text{h.c.}\\
\textbf{E}_{t} &= \vec{\epsilon}_{t}\mathscr{E}_{t}\hat{b}e^{i\vec{k_{t}}\cdot\vec{r}}+\text{h.c.}\\
\end{aligned}
\end{equation}
where we have assumed fields propagate in $z$ direction and
\begin{equation}
\begin{aligned}
\mathscr{E}_{s} &= \left(\frac{\hbar\omega_{s}}{2\epsilon_0V}\right)^{1/2}\\
\mathscr{E}_{t} &= \left(\frac{\hbar\omega_{t}}{2\epsilon_0V}\right)^{1/2}.
\end{aligned}
\end{equation}
Under the rotating wave approximation, we can obtain that
\begin{equation}
\begin{aligned}
\sum_i^N\hat{\textbf{d}}^i\cdot\textbf{E} &= \sum_i^N\left[\textbf{d}_{12}\cdot\vec{\epsilon}_I\frac{\mathscr{E}_I}{2}\sigma^i_{21}e^{-i\omega^L_It+i\vec{k_{I}}\cdot \vec{r}_i}+\textbf{d}_{13}\cdot\vec{\epsilon}_{s}\mathscr{E}_{s}\hat{a}\sigma^i_{31}e^{i\vec{k_{s}}\cdot\vec{r}_i}\right. \\
&\ \ \ \ \ \left.+\textbf{d}_{24}\cdot\vec{\epsilon}_{t}\mathscr{E}_{t}\hat{b}\sigma^i_{42}e^{i\vec{k_{t}}\cdot\vec{r}_i}+\textbf{d}_{34}\cdot\vec{\epsilon}_{II}\frac{\mathscr{E}_{II}}{2}\sigma^i_{43}e^{-i\omega^L_{II}t+i\vec{k_{II}}\cdot \vec{r}_i}+h.c.\right]\\
&=-\sum_i^N[\hbar\frac{\Omega_{I}}{2}\sigma^i_{21}e^{-i\omega^L_It+i\vec{k_{I}}\cdot \vec{r}_i}+\hbar\frac{\Omega_{II}}{2}\sigma^i_{43}e^{-i\omega^L_{II}t+i\vec{k_{II}}\cdot \vec{r}_i}\\
&\ \ \ \ \
+\hbar g_{s}\hat{a}\sigma^i_{31}e^{i\vec{k_{s}}\cdot\vec{r}_i}+\hbar g_{t}\hat{b}\sigma^i_{42}e^{i\vec{k_{t}}\cdot\vec{r}_i}+h.c.],
\end{aligned}
\end{equation}
so the full Hamiltonian is
\begin{equation}
\begin{aligned}
\hat{H} &=\sum^N_{i=1}[\hbar\omega_2\sigma^i_{22}+\hbar\omega_3\sigma^i_{33}+\hbar\omega_4\sigma^i_{44}+(\hbar\frac{\Omega_{I}}{2}\sigma^i_{21}e^{-i\omega^L_It+i\vec{k_{I}}\cdot \vec{r}_i}+\hbar\frac{\Omega_{II}}{2}\sigma^i_{43}e^{-i\omega^L_{II}t+i\vec{k_{II}}\cdot \vec{r}_i}\\
&\ \ \ \ \
+\hbar g_{s}\hat{a}\sigma^i_{31}e^{i\vec{k_{s}}\cdot\vec{r}_i}+\hbar g_{t}\hat{b}\sigma^i_{42}e^{i\vec{k_{t}}\cdot\vec{r}_i}+h.c.)]+\hbar\omega_{a}a^{\dagger}a+\hbar\omega_{b}b^{\dagger}b\\
\end{aligned}
\end{equation}
We first define the unitary transformation of individual atoms as
\begin{equation}
\begin{aligned}
U_R^i &= e^{-i(\omega_{II}+\omega_{s})t+i(\vec{k_{II}}+\vec{k_{s}})\cdot \vec{r}_i}\sigma^i_{44}+e^{-i\omega_{s}t+i\vec{k}_s\cdot\vec{r}_i}\sigma^i_{33}+e^{-i\omega_{I}t+i\vec{k}_I\cdot\vec{r}_i}\sigma^i_{22}+\sigma^i_{11},
\end{aligned}
\end{equation}
from which we define a global unitary transformation,
\begin{equation}
\begin{aligned}
U_R&=U^1_{R,atom}\otimes U^2_{R,atom}...\otimes U^N_{R,atom}\otimes e^{-i(\omega_{s}a^{\dagger}a+\omega_{t}b^{\dagger}b)t},
\end{aligned}
\end{equation}
so a state ket transform as $\ket{\tilde{a}}=U_R^{\dagger}\ket{a}$. Here $\omega_I$ and $\omega_{II}$ are the laser frequency of pump field I and II, while $\omega_s$ and $\omega_t$ are the frequency corresponding to input signal photon $s$ and output telecom photon $t$. The Hamiltonian transforms as
\begin{equation}
\begin{aligned}
\tilde{H} &= U_R^{\dagger}\hat{H}U_R+i\hbar(\partial_t{U_R^{\dagger}}U_R).
\end{aligned}
\end{equation}
Thus, the full Hamiltonian in the rotating frame takes the form
\begin{equation}
\begin{aligned}
\tilde{H} &= \sum^N_{i=1}[\hbar\Delta_2\sigma^i_{22}+\hbar\Delta_3\sigma^i_{33}+\hbar(\Delta_4)\sigma^i_{44}\\
&\ \ \ \ +(\hbar\frac{\Omega_{I}}{2}\sigma^i_{21}+\hbar\frac{\Omega_{II}}{2}\sigma^i_{43}
+\hbar g_{s}\hat{a}\sigma^i_{31}+\hbar g_{t}\hat{b}\sigma^i_{42}+h.c.)]\\
&\ \ \ \ +\hbar\Delta_aa^{\dagger}a+\hbar\Delta_bb^{\dagger}b,
\end{aligned}
\end{equation}
where we assumed energy conservation $\omega_{II}+\omega_s-\omega_I-\omega_t = 0$ and phase match condition $\vec{k_{t}}+\vec{k_{I}}-\vec{k_{II}}-\vec{k_{s}} = 0$. The detunings are defined as $\Delta_2 = \omega_2 - \omega_I$, $\Delta_3 = \omega_3 - \omega_s$, $\Delta_4 = \omega_4 - \omega_{II} - \omega_s$, $\Delta_a = \omega_a - \omega_s$ and $\Delta_b = \omega_b - \omega_t$.  

We introduce atomic collective excitation operators, defined as
\begin{equation}
\begin{aligned}
S_2^{\dagger} & = \frac{1}{\sqrt{N}}\sum_i^N\sigma^i_{21},\\
S_3^{\dagger} & = \frac{1}{\sqrt{N}}\sum_i^N\sigma^i_{31},\\
S_4^{\dagger} & = \frac{1}{\sqrt{N}}\sum_i^N\sigma^i_{41}.
\end{aligned}
\end{equation}
We assume the system is in a weak excitation region, and the atomic states are always in the symmetric collective excitation manifold. This is a reasonable assumption for atomic state $\ket{3}$ and $\ket{4}$, since input 780nm are of few photon level. This is also a reasonable assumption for state $\ket{2}$ due to the decay channels from $\ket{2}$ to the other ground hyperfine states outside of the four-wave mixing loop. With this low excitation assumption, the $S_i$ operators approximately obey the bosonic annihilation and creation operator commutation relation(\textit{55}).
%for example, consider a two level atom system with ground state $\ket{g}$ and $\ket{e}$, the excited states of the   collective symmetric state of an atomic ensemble of atom number N is
%\begin{equation}
%\begin{aligned}
%\ket{\text{Symmetric}} &= \frac{1}{\sqrt{N}}(\ket{eg...g}+\ket{ge...g}+\ket{gge...g}+...+\ket{g...ge}).
%\end{aligned}
%\end{equation}
Within this symmetric manifold, we can also write
\begin{equation}
\begin{aligned}
\sum_i^N\sigma^i_{22} &= S^{\dagger}_2S_2\\
\sum_i^N\sigma^i_{33} &= S^{\dagger}_3S_3\\
\sum_i^N\sigma^i_{44} &= S^{\dagger}_4S_4\\
\sum_i^N\sigma^i_{43} &= S^{\dagger}_4S_3\\
\sum_i^N\sigma^i_{42} &= S^{\dagger}_4S_2.\\
\end{aligned}
\end{equation}
The Hamiltonian is then
\begin{equation}
\begin{aligned}
\tilde{H} &= \hbar [\Delta_2S^{\dagger}_2S_2+\Delta_3S^{\dagger}_3S_3+\Delta_4S^{\dagger}_4S_4\\
&\ \ \ \ +(\frac{\sqrt{N}\Omega_{I}}{2}S_2^{\dagger}+\frac{\Omega_{II}}{2}S^{\dagger}_4S_3
+\sqrt{N}g_{s}\hat{a}S_3^{\dagger}+ g_{t}\hat{b}S^{\dagger}_4S_2+h.c.)]\\
&\ \ \ \ +\hbar\Delta_aa^{\dagger}a+\hbar\Delta_bb^{\dagger}b\\
\end{aligned}
\end{equation}

The Heisenberg-Langevin equation is(\textit{54}, \textit{56})%~\cite{gardiner1985input,gardiner2004quantum}
\begin{equation}
\begin{aligned}
\dot{a} &= (-\frac{\kappa_a}{2}-i\Delta_a)a-i\sqrt{N}g_s S_3-\sqrt{\kappa_a}a_{in}(t)\\
\dot{b} &=(-\frac{\kappa_b}{2}-i\Delta_b)b-ig_t S^{\dagger}_2S_4-\sqrt{\kappa_b}b_{in}(t)\\
\dot{S_2} &= (-\frac{\Gamma_2}{2}-i\Delta_2)S_2-ig_tb^{\dagger}S_4-i\sqrt{N}\frac{\Omega_I}{2}\\
\dot{S_3} &= (-\frac{\Gamma_3}{2}-i\Delta_3)S_3-i\sqrt{N}g_sa-i\frac{\Omega_{II}}{2}S_4\\
\dot{S_4} &= (-\frac{\Gamma_4}{2}-i\Delta_4)S_4-ig_tbS_2-i\frac{\Omega_{II}}{2}S_3\\
\end{aligned}
\end{equation}
where we set input noise terms $S_{i,\ in}(t) = 0$. 
In our experiment, we set pump II detuning $\Delta_2 = 0$. It is then reasonable to find steady state by setting $\dot{S_2} = 0$. Thus we have
\begin{equation}
\begin{aligned}
(-\frac{\Gamma_2}{2}-i\Delta_2)S_2&=ig_tb^{\dagger}S_4+i\sqrt{N}\frac{\Omega_I}{2},
\end{aligned}
\end{equation}
we also drop the small non linear term. We have
\begin{equation}
\begin{aligned}
S_2&=\frac{i\sqrt{N}\frac{\Omega_I}{2}}{-\frac{\Gamma_2}{2}-i\Delta_2}=A.
\end{aligned}
\end{equation}
Thus, all nonlinear terms can be eliminated by substituting $S_2$, we can write the equation in matrix form
\begin{equation}
\begin{aligned}
\begin{pmatrix}
\dot{a}\\
\dot{b}\\
\dot{S}_3\\
\dot{S}_4\\
\end{pmatrix} &= 
\begin{pmatrix}
(-\frac{\kappa_a}{2}-i\Delta_a) & 0 & -i\sqrt{N}g_s & 0\\
0 & (-\frac{\kappa_b}{2}-i\Delta_b) & 0 & -ig_t A^{\dagger}\\
-i\sqrt{N}g_s& 0&(-\frac{\Gamma_3}{2}-i\Delta_3)& -i\frac{\Omega_{II}}{2}\\
0 & -ig_tA& -i\frac{\Omega_{II}}{2}&(-\frac{\Gamma_4}{2}-i\Delta_4)
\end{pmatrix}
\begin{pmatrix}
a\\
b\\
S_3\\
S_4\\
\end{pmatrix}\\
&+\begin{pmatrix}
-\sqrt{\kappa_a}a_{in}\\
-\sqrt{\kappa_b}b_{in}\\
0\\
0\\
\end{pmatrix}
\end{aligned}
\end{equation}
Now we define the Fourier transform of the operators in the rotating frame
\begin{equation}
\begin{aligned}
a(\omega') &= \frac{1}{\sqrt{2\pi}}\int^{\infty}_{-\infty}dt\ a(t)e^{i\omega'(t-t_0)}
\end{aligned}
\end{equation}
%We do not assume $a(\omega')=a_{\omega'}$. We can think of $a(\omega')$ as an operator satisfying
%\begin{equation}
%\begin{aligned}
%a(t) &= \frac{1}{\sqrt{2\pi}}\int^{\infty}_{-\infty}d\omega'\ a(\omega')e^{-i\omega'(t-t_0)},
%\end{aligned}
%\end{equation}
thus in frequency space the above matrix equation becomes
\begin{equation}
\begin{aligned}
&
\begin{pmatrix}
(i\omega-\frac{\kappa_a}{2}-i\Delta_a) & 0 & -i\sqrt{N}g_s & 0\\
0 & (i\omega-\frac{\kappa_b}{2}-i\Delta_b) & 0 & -ig_t A^{\dagger}\\
-i\sqrt{N}g_s& 0&(i\omega-\frac{\Gamma_3}{2}-i\Delta_3)& -i\frac{\Omega_{II}}{2}\\
0 & -ig_tA& -i\frac{\Omega_{II}}{2}&(i\omega-\frac{\Gamma_4}{2}-i\Delta_4)
\end{pmatrix}
\begin{pmatrix}
a(\omega)\\
b(\omega)\\
S_3(\omega)\\
S_4(\omega)\\
\end{pmatrix}\\
&=\begin{pmatrix}
\sqrt{\kappa_a}a_{in}\\
\sqrt{\kappa_b}b_{in}\\
0\\
0\\
\end{pmatrix}
\end{aligned}
\end{equation}
Solving the linear equation, we obtain
\begin{equation}
\begin{aligned}
b(\omega) &= \frac{-(-ig_t A^{\dagger})(-i\frac{\Omega_{II}}{2})(-i\sqrt{N}g_s)\sqrt{\kappa_a}}{A(\omega)-B(\omega)-C(\omega)-D(\omega)+E(\omega)}a_{in}(\omega),
\end{aligned}
\end{equation}
where
\begin{equation}
\begin{aligned}
A(\omega) &= (i\omega-\frac{\kappa_a}{2}-i\Delta_a)(i\omega-\frac{\kappa_b}{2}-i\Delta_b)(i\omega-\frac{\Gamma_3}{2}-i\Delta_3)(i\omega-\frac{\Gamma_4}{2}-i\Delta_4),\\
B(\omega) &= (i\omega-\frac{\kappa_a}{2}-i\Delta_a)(i\omega-\frac{\Gamma_3}{2}-i\Delta_3)(-ig_t \frac{-i\sqrt{N}\frac{\Omega_I}{2}}{-\frac{\Gamma_2}{2}+i\Delta_2})(-ig_t\frac{i\sqrt{N}\frac{\Omega_I}{2}}{-\frac{\Gamma_2}{2}-i\Delta_2}),\\
C(\omega) &=(i\omega-\frac{\kappa_a}{2}-i\Delta_a)(i\omega-\frac{\kappa_b}{2}-i\Delta_b)(-i\frac{\Omega_{II}}{2})^2,\\
D(\omega) &= (i\omega-\frac{\kappa_b}{2}-i\Delta_b)(i\omega-\frac{\Gamma_4}{2}-i\Delta_4)(-i\sqrt{N}g_s)^2,\\
E(\omega) &= (-ig_t \frac{-i\sqrt{N}\frac{\Omega_I}{2}}{-\frac{\Gamma_2}{2}+i\Delta_2})(-ig_t\frac{i\sqrt{N}\frac{\Omega_I}{2}}{-\frac{\Gamma_2}{2}-i\Delta_2})(-i\sqrt{N}g_s)^2.
\end{aligned}
\end{equation}
In the input-output formalism, the input-output relation is
\begin{equation}
\begin{aligned}
b(\omega) & = [b_{out}(\omega)-b_{in}(\omega)]/\sqrt{\kappa_b}.
\end{aligned}
\end{equation}
In the conversion system, we only input 780nm as mode $a_{in}(\omega)$ so we can set $b_{in}(\omega)=0$ , the input output relation becomes
\begin{equation}
\begin{aligned}
b_{out}(\omega) &= \frac{-(-ig_t A^{\dagger})(-i\frac{\Omega_{II}}{2})(-i\sqrt{N}g_s)\sqrt{\kappa_a\kappa_b}}{A(\omega)-B(\omega)-C(\omega)-D(\omega)+E(\omega)}a_{in}(\omega).
\end{aligned}
\end{equation}
We define
\begin{equation}
\begin{aligned}
\eta(\omega) &= \frac{-(-ig_t A^{\dagger})(-i\frac{\Omega_{II}}{2})(-i\sqrt{N}g_s)\sqrt{\kappa_a\kappa_b}}{A(\omega)-B(\omega)-C(\omega)-D(\omega)+E(\omega)},
\end{aligned}
\end{equation}
such that
\begin{equation}
\begin{aligned}
b_{out}^{\dagger}(\omega)b_{out}(\omega) &= |\eta(\omega)|^2\  a_{in}^{\dagger}(\omega)a_{in}(\omega).
\end{aligned}
\end{equation}
We will calculate the photon flux operator $b_{out}^{\dagger}(t)b_{out}(t)$ , from which we can obtain the mean photon flux of the output 1324nm signal. By definition
\begin{equation}
\begin{aligned}
b_{out}(t) &= \frac{i}{\sqrt{2\pi}}\int^{\infty}_{-\infty}d\omega'\ b_1(\omega')e^{-i\omega'(t-t_1)},
\end{aligned}
\end{equation}
where $b_1(\omega)$ is understand to be field operator $b_{\omega}$ outside the cavity at $t=t_1$ in the future time.
\begin{equation}
\begin{aligned}
a_{in}(t) &= \frac{i}{\sqrt{2\pi}}\int^{\infty}_{-\infty}d\omega'\ a_0(\omega')e^{-i\omega'(t-t_0)},
\end{aligned}
\end{equation}
where $a_0(\omega)$ is understand to be field operator $a_{\omega}$ outside the cavity at $t=0$.
Comparing the definition and the operator Fourier transform, we have
\begin{equation}
\begin{aligned}
a_{in}(\omega)&=ia_0(\omega)\\
b_{out}(\omega)&=ie^{i\omega(t_1-t_0)}b_1(\omega),
\end{aligned}
\end{equation}
however, we usually use the annihilation and creation operator in normal order, they always appear in conjugate pairs, so
\begin{equation}
\begin{aligned}
a^{\dagger}_{in}(\omega)a_{in}(\omega)&=a^{\dagger}_0(\omega)a_0(\omega)\\
b^{\dagger}_{out}(\omega)b_{out}(\omega)&=b^{\dagger}_1(\omega)b_1(\omega).
\end{aligned}
\end{equation}
Also, by definition of the Fourier transform,
\begin{equation}
\begin{aligned}
b_{out}(t) &= \frac{1}{\sqrt{2\pi}}\int^{\infty}_{-\infty}d\omega'\ b_{out}(\omega')e^{-i\omega'(t-t_0)},
\end{aligned}
\end{equation}
so we can write
\begin{equation}
\begin{aligned}
\langle b_{out}^{\dagger}(t)b_{out}(t)\rangle &=\frac{1}{2\pi}\int^{\infty}_{-\infty}\int^{\infty}_{-\infty}d\omega_1d\omega_2\ e^{-i(\omega_2-\omega_1)(t-t_0)}\eta^*(\omega_1)\eta(\omega_2)\langle a_{0}^{\dagger}(\omega_1)a_{0}(\omega_2) \rangle\\
\end{aligned}
\end{equation}
The LHS is exactly the mean photon flux at time $t$. To find out conversion efficiency, in our case we note that contribution of  $\langle a_{0}^{\dagger}(\omega_1)a_{0}(\omega_2)\rangle$ only comes from near center frequency $\omega_0$ , so we make the narrow-bandwidth approximation that
\begin{equation}
\begin{aligned}
\langle b_{out}^{\dagger}(t)b_{out}(t)\rangle &= \eta^*(\omega_0)\eta(\omega_0)\frac{1}{2\pi}\int^{\infty}_{-\infty}\int^{\infty}_{-\infty}d\omega_1d\omega_2\ e^{-i(\omega_2-\omega_1)(t-t_0)}\langle a_{0}^{\dagger}(\omega_1)a_{0}(\omega_2) \rangle\\
&=|\eta(\omega_0)|^2\langle a_{in}^{\dagger}(t)a_{in}(t)\rangle,
\end{aligned}
\end{equation}
thus the conversion efficiency is $|\eta(\omega_0)|^2$. In the weak excitation region, this conversion efficiency is independent of the input photon number. In our experimental setup, $\Delta_2=\Delta_3=\Delta_4=0$, and $\Delta_a=\Delta_b=0$. In the limit of free-space coupling, i.e. very broadband cavity (large $\kappa_a$ and large $\kappa_b$), we can approximate the conversion efficiency as  

\begin{equation}
    |\eta(0)|^2\approx\left|\frac{4Ng_s\,g_t\,\Omega_I\,\Omega_{II}}{\Gamma_2\left(\Omega_{II}^2+\Gamma_3\,\Gamma_4/2\right)}\right|^2\frac{1}{\kappa_a\kappa_b}.
\end{equation}

\section{HOM Interference Model}

\subsection{Setup and notation}

We are considering the situation pictured in Figure 7, where two independent coherent beams are incident on the input ports $a$ and $b$ of a standard (ie ideal 50:50, non-polarizing) beam splitter; and we want to calculate the rate of coincidences observed at the two output ports $c$ and $d$ as a function of two specific measurement times $t_{c}$ and $t_{d}$.  Our main goal is to calculate what the shape of this coincidence rate looks like versus the arrival time difference $\tau \equiv t_{c}-t_{d}$.  In the case of CW beams this will show us the HOM interference dip, and we will then consider the case of the beams being pulsed in time.

\paragraph{Initial beam states. }We assume the incoming beams each come from a laser with a small but finite range in linewidth.  Each individual mode $k$ within the beam is then assumed to be in a coherent state $\ket{\alpha}_{k}$, where: 
\begin{equation}
\hat{a}_{k} \ket{\alpha}_{k} = \alpha \ket{\alpha}_{k}
\label{eq:basic_coherence}
\end{equation}

\noindent
 In an finite quantization volume, we denote $\alpha$ corresponds to a single mode as $\alpha(\nu_{k})=\alpha_{k}$. We further expand the definition to continuous frequency space by making quantization volume infinite and use the notation $\alpha(\nu)$. 

If we assign $\alpha$ to describe the beam entering at port $a$, and a corresponding spectral function $\beta(\nu)$ for the beam at port $b$, as in Figure 7, then we can write the complete incoming state as
\begin{equation}
   \ket{\Psi}_{a,b} = \ket{\{ \alpha \}}_{a} \otimes 
   \ket{\{\beta\}}_{b} 
   \label{eq:in_state}
\end{equation}

\paragraph{Intensity, transforms and coherence time: Weiner-Khinchin interlude. }Following the general Glauber theory of photo-detection we can write the intensity in any beam for a field with state $\ket{\Psi}$ at any given time $t$ as

\begin{equation}
    I(t) \equiv
    \frac{dN}{dt}=\int \frac{dN}{d\nu \, dt} d\nu \, \propto \, 
    \bra{\Psi} \hat{\mathcal{E}}^{(-)} (t) \, \hat{\mathcal{E}}^{(+)} (t) \, \ket{\Psi} 
     =  \left| \, \hat{\mathcal{E}}^{(+)} (t) \, \ket{\Psi} \right|^2
    \label{eq:basic_Glauber}
\end{equation}

\noindent
Here $\hat{\mathcal{E}}^{(+)} (t)$ is the positive-frequency electric field operator:
\begin{equation}
    \hat{\mathcal{E}}^{(+)} (t) \; \propto \sum_{k} 
    \hat{a}_{k} \; e^{+i \, \nu_{k} \, t \, 2\pi} 
    \propto \int \hat{a}_{\nu} \, e^{+i \nu t 2\pi}  d\nu
    \label{eq:Eplus_def}
\end{equation}
\noindent
Applying this to our beam with state $\ket{\Psi}=\ket{\{\alpha\}}$:

\begin{equation}
        I(t)  \propto  \left| \, \sum_{k} 
    \alpha(\nu_{k}) \; e^{+i \, \nu_{k} \, t \, 2\pi} \, \right|^2  
           \propto  \left| \, \int \alpha(\nu) \, e^{+i \nu t 2\pi}  d\nu
         \, \right|^2
         \label{eq:basic_intensity}
\end{equation}
\noindent
Define the Fourier transform of $\alpha(\nu)$, which we denote as $A(t)$
\begin{equation}
    A(t) \equiv \mathcal{F}^{-1}\left[ \alpha(\nu) \right](t) \equiv
    \int \alpha(\nu) \, e^{+i \nu t 2\pi}  d\nu 
    \quad \text{and so} \quad
    I(t) \propto \left| A(t) \right|^2
    \label{eq:FT_A}
\end{equation}
\noindent
as is also shown in channel $a$ of Figure 7.  We will use Equation~\ref{eq:FT_A} to define our convention for an inverse Fourier transform, ie from frequency to time, with the forward transform then being
\begin{equation}
    \alpha(\nu) = \mathcal{F}\left[ A(t) \right](\nu) \equiv 
    \int A(t) \, e^{-i \nu t 2\pi}  dt
    \label{eq:FT_alpha}
\end{equation}
\noindent
and so refer to the two as a Fourier transform pair, denoted
\begin{equation}
    A(t) \stackrel{\mathcal{FT}}{\leftrightarrow}  \alpha(\nu).
\end{equation}

\medskip

\noindent
We also use the notation for auto-correlation:
\begin{equation}
    \text{Autocorrelation}\left[ A(t)\right](\Delta t) \equiv 
    (A \star A)(\Delta t) \equiv
    \int A(t'+\Delta t) A^{*}(t') dt'
\end{equation}
\noindent

\subsection{Pair Rates and Coincidence Distribution}

Looking back now to Figure~7 we describe the input beam at point $a$ with the spectral function $\alpha(\nu)$, which has transform $A(t)$, and the input beam at point $b$ has spectral function $\beta(\nu)$, and corresponding transform $B(t)$.  After writing the output state, we will calculate the rates for observing pairs at the outputs, and then consider both the case of CW beams and envelope-pulsed beams.

%%%%%%%%%%%%%%%%%%%%%%%%%%%%%%
% Subsection on beams splitter effect
%
%\subsubsection{Output state and pair detection}
%In general, higher Fock states do not pass smoothly through beam splitters.  A perfectly simple, well-defined energy eigenstate with occupation number $n$ will result in an output state that is a superposition of some number $m$ photons having come out one side and $k$ out the other, for all possible combinations of $m+k=n$.  As a general procedure it will be simpler, then, to propagate the operators at the output locations back to the input locations, than it is to move the input state to the output state.

%However, coherent states are very much the exception to this rule.  As might be expected from the counterpart to classical EM waves a coherent state at a beamsplitter input is simply transformed into two coherent states, one at each output.  Specifically, f
For a coherent state defined by $\ket{\{ \alpha \}}_{\mathrm{Input}}$ at the input to a 50:50 symmetric beamsplitter the state of the two outputs is $\ket{\{ \alpha/\sqrt{2} \}}_{\mathrm{Out1}}  \otimes \ket{\{ \alpha/\sqrt{2} \}}_{\mathrm{Out2}}$.  We can see that this evolution conserves energy while both beams inherit the same phase relation between modes as in the original beam. With two input beams we can simply add their amplitudes at the two outputs, keeping in mind that one of them has to be phase-reversed (this is needed to conserve energy and so a general property of beam splitters). The states of the output beams at port locations $c$ and $d$, and their transforms, are 
\begin{eqnarray}
   & & \ket{\Psi} = \ket{\{(\alpha+\beta)/\sqrt{2}\}}_{c} \otimes \ket{\{(\alpha-\beta)/\sqrt{2}\}}_{d} \label{eq:out_state} \\
    & & \left( A(t) + B(t) \right)/\sqrt{2} \quad \text{at c and} \,\, \left( A(t) - B(t) \right)/\sqrt{2} \quad \text{at d}  \label{eq:out_waves}
\end{eqnarray}

With this we can write the rate for two-photon observations at ports $c$ and $d$, specifically at times $t_{c}$ and $t_{d}$ as
\begin{equation}
    \frac{d^{2}N^{(2)}}{dt_{c} \; dt_{d}}  \propto  \left|  
    \hat{\mathcal{E}}^{(+)_{c}}(t_{c}) \;  \hat{\mathcal{E}}^{(+)_{d}}(t_{d})
    \ket{\Psi} \right|^2 \\
    \label{eq:pair_rate_basic}
\end{equation}
\noindent
Substituting Equations~\ref{eq:Eplus_def} and~\ref{eq:out_state} into the above then yields
\begin{eqnarray}
\frac{d^{2}N^{(2)}}{dt_{c} \; dt_{d}}  & \propto &  \frac{1}{4} 
\left|  
 \int (\alpha(\nu')+\beta(\nu')) e^{+i \, \nu' t_{c} \, 2\pi} d\nu' 
 \int (\alpha(\nu'')-\beta(\nu'')) e^{+i \, \nu'' t_{d} \, 2\pi} d\nu''
    \; \right|^2 \nonumber \\
  & = & \frac{1}{4} 
  \left|  
  \left( A(t_{c})+B(t_{c}) \right) \left( A(t_{d})-B(t_{d}) \right) 
  \right|^{2}  
  \label{eq:d2Ndtcdtd}
\end{eqnarray}
\noindent
Expanding Eq.~\ref{eq:d2Ndtcdtd} gives us three types of terms:
\begin{itemize}
    \item Balanced, synchronous terms which contain only combinations of products $A() \, A^{*}()$ and/or $B() \, B^{*}()$ with both paired factors evaluated at the same time.  These terms are immediately real and correspond to products of the beam intensities, see below.

    \item Balanced, asynchronous terms containing $A() \, A^{*}()$ and $B() \, B^{*}()$ products which are evaluated at differing times.  These terms are where the interference effects will stem from.

    \item Unbalanced terms, where at least one factor of $A$ or $B$ is not balanced by its complex conjugate.  These terms will have a very fast varying phase at all times and will vanish in any time integration such as we are about to carry out in Equation~\ref{eq:pair_rate_Dt_1} below.
\end{itemize}

\noindent
If we drop the unbalanced terms then Eq.~\ref{eq:d2Ndtcdtd} can be re-arranged to
\begin{eqnarray}
    \frac{d^{2}N^{(2)}}{dt_{c} \; dt_{d}} & \propto &
    \frac{1}{4} \left[ I_{a}(t_{c}) I_{a}(t_{d}) + I_{b}(t_{c}) I_{b}(t_{d}) +
    I_{a}(t_{c}) I_{b}(t_{d}) + I_{b}(t_{c}) I_{a}(t_{d})
    \right] \nonumber \\
    & & \quad - \frac{1}{2} \; \textbf{Re} \left\{ A(t_{c}) B^{*}(t_{c}) 
       B(t_{d}) A^{*}(t_{d}) \right\}
    \label{eq:pair_rate_from_AB}
\end{eqnarray}
\noindent
where we have identified the single-photon intensities in the two incoming beams very naturally as $I_{a}(t)\equiv|A(t)|^2$ and $I_{b}(t)\equiv|B(t)|^2$. 

%\medskip
%The result of Equation~\ref{eq:pair_rate_from_AB} admits a simple and pleasing interpretation, namely: (i)~The first line of the RHS is the ``combinatoric'' pair rate, corresponding to the purely classical product of the probabilities of getting individual photons from the single beams; the first two terms are the rates for two photons from one beam, the latter one photon from each beam; and (ii)~The second line is then the ``interference'' term embodying the HOM effect and always involves both beams; and, note that the intereference term is strictly negative.  Further, a little thought will show that the interference term will only be significant if the time difference $\Delta t = t_{d} - t_{c}$ is on the order of, or smaller than, the beams' coherence time.

\smallskip
For the experiment we integrate the number of observed pairs over time, binned on the time difference between the detections.  Accordingly we will change variables from $[ t_{c},t_{d} ]$ to $[ \, t_{c}, \, \Delta t \equiv t_{d}-t_{c} ]$ and then integrate the pair rate over all $t_{c}$
\begin{eqnarray}    
    \frac{dN^{(2)}}{d \, \Delta t} & = & 
    \int \frac{d^{2}N^{(2)}}{dt_{c} \; d \Delta t} dt_{c} \\
    & \propto &
    \frac{1}{4} \int \left[ I_{a}(t_{c}) I_{a}(t_{c}+\Delta t) + 
    I_{b}(t_{c}) I_{b}(t_{c}+\Delta t) +
    I_{a}(t_{c}) I_{b}(t_{c}+\Delta t) + 
    I_{b}(t_{c}) I_{a}(t_{c}+\Delta t) \right] dt_{c} \nonumber \\
    & & \quad - \frac{1}{2} \int
     \; \textbf{Re} \left\{ A(t_{c}) B^{*}(t_{c}) 
       B(t_{c}+\Delta t) A^{*}(t_{c}+\Delta t) \right\} \, dt_{c}
       \label{eq:pair_rate_Dt_1}
\end{eqnarray}
\noindent
We can recognize the integral of each of the product terms in Equation~\ref{eq:pair_rate_Dt_1} as having the form of either an auto-correlation or cross-correlation.  We can exchange the order of integration and the $\textbf{Re}\{\}$ operation this then simplifies to the pleasingly compact form in terms of just two autocorrelations:
\begin{eqnarray}
    \frac{dN^{(2)}}{d \, \Delta t} & \propto & \left[ 
    (I_{a} \star I_{a})(\Delta t) +
    (I_{b} \star I_{b})(\Delta t) +
    (I_{a} \star I_{b})(\Delta t) +
    (I_{b} \star I_{a})(\Delta t)    
    \right] \nonumber \\
    & & \quad - \frac{1}{2} \textbf{Re} \left\{ 
    \int  A(t_{c}) B^{*}(t_{c}) 
       B(t_{c}+\Delta t) A^{*}(t_{c}+\Delta t) \, dt_{c}
    \right\} \nonumber \\
    & = & \frac{1}{4} ((I_{a}+I_{b}) \star (I_{a}+I_{b}))(\Delta t) 
    - \frac{1}{2} \textbf{Re} \left\{ 
    ((A \, B^{*}) \star (A \, B^{*}))(\Delta t)
    \right\}  \label{eq:pair_rate_Dt_autocorr}
\end{eqnarray}

\smallskip
The form of Equation~\ref{eq:pair_rate_Dt_autocorr} is not fully practical yet: the first, combinatoric term follows just from the beams' power envelopes, but the second, interference term is not yet connected to measurable properties of the individual beams.  To make further progress we need to specify more about the those properties.  We will look at the case for CW beams in Section~\ref{subsec:dip_CW} next, and then the more general case of envelope-pulsed beams in Section\ref{subsec:extension_to_pulsed_case} that follows.

%%%%%%%%%%%%%%%%%%%%%%%%%%%%%%
% Subsection on interference derived in CW case
%
\subsection{Interference dip, CW case}
\label{subsec:dip_CW}
We start with CW case, where we assume the frequency spread of the beam is relatively small and their spectral distributions are smooth.  In this case the intensities $I_{a},I_{b}$ will be constant and the only structure in the pair rate vs $\Delta t$ will come from the interference term.  The only physical parameters that define the two beams are their lineshapes: for a narrow-band beam the different frequency components will fall out of phase with one another, making the sum phase at any given time effectively random.  Thus we expect to see that the interference term can be expressed purely in terms of the two beam's spectral density functions, or lineshape functions, which we know to be $|\alpha(\nu)|^2$ and $|\beta(\nu)|^2$.

\smallskip
We first re-cast the interference term using $\alpha$ and $\beta$ instead of $A$ and $B$.  Concentrating on just the argument of the $\textbf{Re}\{\}$ operator in Equation~\ref{eq:pair_rate_Dt_autocorr}, after a bit of prestidigitation we can arrive at
\begin{equation}
    ((A \, B^{*}) \star (A \, B^{*}))(\Delta t) = 
    \mathcal{F}^{-1} \left[ \, \left| (\alpha \star \beta)(\nu) \right|^{2} \right]
    \label{eq:HOM_term_recast}
\end{equation}
\noindent
%(We leave it as an exercise for the diligent reader to derive this result; one can go the long way following direct substitution, or pull a trick by using the Weiner-Khinichin theorem again first.)
%Equation~\ref{eq:HOM_term_recast} looks very promising, since we have a convolution of the two spectral functions as the central ingredient.  But this is still not operational, since we only know the amplitude structure of $\alpha(\nu)$ and $\beta(\nu)$ and not their phase structure.  To highlight this let's call out the phase and amplitudes separately for each:
where
\begin{equation}
    \alpha(\nu) \equiv |\alpha(\nu)| \, e^{i \phi_{\alpha}(\nu)} 
    \quad \text{and} \quad
    \beta(\nu) \equiv |\beta(\nu)| \, e^{i \phi_{\beta}(\nu)} 
    \label{eq:alpha_beta_phase}
\end{equation}
\noindent
Now we can re-write the core of Equation~\ref{eq:HOM_term_recast} vis
\begin{eqnarray}
    \left| (\alpha \star \beta)(\nu) \right|^{2} & = &
    \left| \int \alpha(\nu') \beta^{*}(\nu'+\nu) d\nu' \right|^{2} \nonumber \\
    & = & \int \alpha(\nu') \beta^{*}(\nu'+\nu) d\nu' 
        \int \alpha^{*}(\nu'') \beta(\nu''+\nu) d\nu'' \nonumber \\
    & = & \int \! \int |\alpha(\nu')| \,  |\alpha(\nu'')| \,
    |\beta(\nu'+\nu)| \,  |\beta(\nu''+\nu)| \nonumber \\
    & & \qquad \times \; 
      \exp{i(\phi_{\alpha}(\nu') - \phi_{\alpha}(\nu''))} \nonumber \\
    & & \qquad \times \;
      \exp{i(\phi_{\beta}(\nu'+\nu) - \phi_{\beta}(\nu''+\nu))}
    \; d\nu' d\nu''
    \label{eq:HOM_core_with_exp}
\end{eqnarray}

%Now we will make a short, qualitative argument, which is nonetheless essentially correct and will bring us over the finish line.  
Since we expect the phase functions $\phi_{\alpha}(\nu)$ and $\phi_{\beta}(\nu)$ to vary very quickly, and almost randomly, as a function of $\nu$; basically, changing $\nu$ by an amount on the order of the inverse of the quantization time will bring a new and unrelated value to the phase.  From this, then, we can see that the integral of Equation~\ref{eq:HOM_core_with_exp} will have essentially no contribution due to the rapidly and randomly shifting complex phases, {\em except} along the line where $\nu'=\nu''$ and the phase differences vanish and the integrand is entirely real.  This then enables us to effectively replace the phase exponential factor with a delta function between $\nu'$ and $\nu''$
\begin{eqnarray}
     \left| (\alpha \star \beta)(\nu) \right|^{2}
    & = & \int \! \int |\alpha(\nu')| \,  |\alpha(\nu'')| \,
    |\beta(\nu'+\nu)| \,  |\beta(\nu''+\nu)| 
    \, \delta(\nu'-\nu'')
    \; d\nu' d\nu''  \nonumber \\
    & = & \int |\alpha(\nu')| \,  |\alpha(\nu')| \,
    |\beta(\nu'+\nu)| \,  |\beta(\nu'+\nu)| 
    \; d\nu'  \nonumber \\
    & = & \int |\alpha(\nu')|^2 \,  |\beta(\nu'+\nu)|^2 
    \; d\nu'  \nonumber \\ 
    & = & (|\alpha|^{2} \star |\beta|^{2})(\nu)
    \label{eq:HOM_core_final}
\end{eqnarray}
With this in hand, then, we can subsitute Equation~\ref{eq:HOM_core_final} into Eq.~\ref{eq:HOM_term_recast} and then into Eq.~\ref{eq:pair_rate_Dt_autocorr} to finally arrive at the pair rate versus time difference as
\begin{equation}
    \frac{dN^{(2)}}{d \, \Delta t}  \; \propto \;
    \frac{1}{4} ((I_{a}+I_{b}) \star (I_{a}+I_{b}))(\Delta t) 
    - \frac{1}{2} \textbf{Re} \left\{ 
    \mathcal{F}^{-1} \left[ \, (|\alpha|^2 \star |\beta|^2)(\nu) 
    \right](\Delta t) \right\}
    \label{eq:pair_rate_Dt_CW_final}
\end{equation}
%
%At first glance this expression may look complicated, with a cross-correlation inside a Fourier transform inside a $\textbf{Re}\{\}$ operation.  But it is well-defined and straightforward to calculate, and we will see in the next Section that it embodies all the behaviors we know to expect from an HOM interference dip feature.

%%%%%%%%%%%%%%%%%%%%%%%%%%%%%%
% Illustrations, starting with the CW case
%
%\subsection{Worked example for CW case}
%\label{subsec:worked_examples_CW}

\paragraph{Gaussian beam lineshapes. }\label{subsubsec:Gaussian_CW_example}We use the notation:
\begin{eqnarray}
    G(t; t_{0}, \sigma_{t}) & \equiv &
    e^{-(t-t_{0})^{2}/2 {\sigma_{t}}^{2}}  \nonumber \\
    g(\nu; \nu_{0}, \sigma_{\nu}) & \equiv &
    e^{-(\nu-\nu_{0})^{2}/2 {\sigma_{\nu}}^{2}}
    \label{eq:Gaussian_symbols}
\end{eqnarray}
Here we continue the convention of using upper-case variables for functions of time and lower-case for functions of frequency.  Note that the functions defined in Equation~\ref{eq:Gaussian_symbols} are not area-normalized, but rather fixed to have a maximum value of 1 at $t=t_{0}$ and $\nu=\nu_{0}$.

%Looking at Equation~\ref{eq:pair_rate_Dt_CW_final} we want to be able to apply two operations to these functions, i.e. cross-correlation and Fourier transform.  Fortunately these are both very straightforward in the case of Gaussians, in that the result is always another Gaussian (plus some complex phase). 

First for the cross-correlation we have:

\begin{eqnarray}
    G(t;t_{1},\sigma_{t,1}) \star G(t;t_{2},\sigma_{t,2}) & = &
    \frac{\sqrt{2 \pi}}{(1/\sigma_{t,1}) \oplus (1/\sigma_{t,2})} \,
    G(t;t_{1}-t_{2},\sigma_{t,1} \oplus \sigma_{t,2}) \nonumber \\
    g(\nu;\nu_{1},\sigma_{\nu,1}) \star g(\nu;\nu_{2},\sigma_{\nu,2}) & = &
    \frac{\sqrt{2 \pi}}{(1/\sigma_{\nu,1}) \oplus (1/\sigma_{\nu,2})} \,
    g(\nu;\nu_{1}-\nu_{2},\sigma_{\nu,1} \oplus \sigma_{\nu,2})
    \label{eq:Gaussian_correlations}
\end{eqnarray}
where we have used the standard notation ``$\oplus$'' for the quadrature sum $x \oplus y = \sqrt{x^2 + y^2}$.  %The constant factor out front is a consequence of the peak normalization choice and not particularly meaningful; we could even safely ignore it and normalize later, but we'll keep everything in for completeness.
Using our convention for the transforms between time $t$ and frequency $\nu$ as laid out in Equations~\ref{eq:FT_A} and~\ref{eq:FT_alpha} we then have:

\begin{eqnarray}
    \mathcal{F}^{-1}\left[ g(\nu; \nu_{0}, \sigma_{\nu}) \right](t) & = &
    \sigma_{\nu} \sqrt{2 \pi} \,
    G(t; 0, 1/2\pi\sigma_{\nu}) \, e^{+i \nu_{0} t 2 \pi} \nonumber \\
    \mathcal{F}\left[ G(t; t_{0}, \sigma_{t}) \right](\nu) & = &
    \sigma_{t} \sqrt{2 \pi} \,
    g(\nu; 0, 1/2\pi\sigma_{t}) \, e^{-i \nu t_{0} 2 \pi}
    \label{eq:Gaussian_transforms}
\end{eqnarray}
where $\sigma_{t} \, \sigma_{\nu} = 1/2\pi$.

For the CW case we have $I_{a}(t)$ and $I_{b}(t)$ constant, and so the auto-correlation in the first term of Equation~\ref{eq:pair_rate_Dt_CW_final} reduces to a simple product, leaving
\begin{equation}
    \frac{dN^{(2)}}{d \, \Delta t}  \; \propto \;
    \frac{1}{4} (I_{a}+I_{b})^{2}
    - \frac{1}{2} \textbf{Re} \left\{ 
    \mathcal{F}^{-1} \left[ \, (|\alpha|^2 \star |\beta|^2)(\nu) 
    \right](\Delta t) \right\}
    \label{eq:pair_rate_Dt_CW_reduced}
\end{equation}

We assume two beams have identical linewidths $\sigma_{\nu}=\sigma_{\nu,1}=\sigma_{\nu,2}$, but might have slightly different central frequencies $\nu_{2}=\nu_{0}$ and $\nu_{1}=\nu_{0}+\delta\nu$.  Then we can write the intensity profiles for the beams at $a$ and $b$ as 
\begin{equation}
    |\alpha|^2(\nu) = \frac{I_{a}}{\sigma_{\nu} \sqrt{2 \pi}} \; g(\nu; \nu_{0}+\delta\nu, \sigma_{\nu}) \quad \mathrm{and} \quad 
    |\beta|^2(\nu) = \frac{I_{b}}{\sigma_{\nu} \sqrt{2 \pi}} \; g(\nu; \nu_{0}, \sigma_{\nu}) 
    \label{eq:alpha_beta_Gaussians}
\end{equation}

Substituting these into Equation~\ref{eq:pair_rate_Dt_CW_reduced} yields
\begin{equation}
    \frac{dN^{(2)}}{d \, \Delta t}  \; \propto \;
    \frac{1}{4} (I_{a}+I_{b})^{2}
    - \frac{1}{2} \, I_{a} I_{b} \;
    G(\Delta t; 0, 1/(\sqrt{2} \, 2 \pi \sigma_{\nu})) \;
    \cos{(2 \pi \, \delta \nu \, \Delta t)}
    \label{eq:pair_rate_Dt_CW_Gaussian}
\end{equation}
for the pair rate versus $\Delta t$.  We can recover the form in terms of a visibility $\mathcal{V}$ if we normalize the rate to the plateau at large $\Delta t$, arriving at 
\begin{equation}
    \frac{dN^{(2)}}{d \, \Delta t}  \; \propto \;
    1
    - \mathcal{V} \;
    G(\Delta t; 0, 1/(\sqrt{2} \, 2 \pi \sigma_{\nu})) \;
    \cos{(2 \pi \, \Delta t \, \delta \nu)} 
    \quad \mathrm{with} \quad 
    \mathcal{V}  \equiv  \frac{2 I_{a} I_{b}}{(I_{a}+I_{b})^2}.
    \label{eq:pair_rate_Dt_CW_Gaussian_visnorm}
\end{equation}
%As usual the visibility reaches its maximum value of 0.5 in the case that the beam intensities are balanced, i.e. $I_{a}=I_{b}$. 

%%%%%%%%%%%%%%%%%%%%%%%%%%%%%%
% Subsection generalizing to pulsed beam case
%
\subsection{Pulsed beam case}
\label{subsec:extension_to_pulsed_case}
%One might be inclined to object at this point, that we seem to have done a lot of work and only -- so far -- gotten a basic fitting form with a two-parameter Gaussian and cosine wiggle.  The bigger payoff, however, comes once we move from the case of CW beams to pulsed beams.  With one more tool, which is to understand the effects of pulsing in terms of frequency spectrum, the rest is already in place in Equation~\ref{eq:pair_rate_Dt_CW_final}.
%\smallskip
%The first or combinatoric term in Equation~\ref{eq:pair_rate_Dt_CW_final} represents the detection of photons appearing indepndently from the two beams.  In the pulsed beam case this term will now have a time structure following from the two intensity envelope functions $I_{a}(t)$ and $I_{b}(t)$ as shown.  To evaluate the second, interference, term, though, we will have to represent the pulsed beam at the level of the electric field amplitude functions $A(t)$ and $B(t)$.

We model the pulsed beam by taking the electric field function for a CW beam and modulating it with a pulse profile function. 

We first represent two ``parent'' CW beams with the electric field functions $X(t)$ and $Y(t)$, having the same properties as the $A$ and $B$ used in the general CW case of Section~\ref{subsec:dip_CW}, and particularly as described in Equation~\ref{eq:alpha_beta_phase}.    Continuing the convention of using upper case for functions of time and corresponding lower case for functions of frequency
\begin{equation}
    X(t) \stackrel{\mathcal{FT}}{\leftrightarrow}  x(\nu) \quad \mathrm{and} \quad
    Y(t) \stackrel{\mathcal{FT}}{\leftrightarrow}  y(\nu)
    \label{eq:X_Y_x_y}
\end{equation}
and then
\begin{equation}
    x(\nu) = |x(\nu)| \, e^{i \phi_{x}(\nu)} 
    \quad \text{and} \quad
    y(\nu) = |y(\nu)| \, e^{i \phi_{y}(\nu)} 
    \label{eq:x_y_phase}
\end{equation}
with $\phi_{x}(\nu)$ and $\phi_{y}(\nu)$ being fast-varying, essentially random functions of $\nu$.

Now we define two pulse envelope amplitude functions $P(t)$ and $Q(t)$ for the two beams, making the full electric field functions at the two inputs
\begin{eqnarray}
    A(t) = X(t) \, P(t) & \quad & B(t) = Y(t) \, Q(t)  
    \label{eq:pulsed_A_B} \\
    I_{a}(t) = |X(t)  P(t)|^{2} & \quad &
    I_{b}(t) = |Y(t)  Q(t)|^{2}
    \label{eq:pulsed_Ia_Ib}
\end{eqnarray}
Note that, even for given CW beams $X$ and $Y$, the pulse envelope functions $P$ and $Q$ which will produce a given $I_{a}$ and $I_{b}$ as in Equation~\ref{eq:pulsed_Ia_Ib} are not unique, but could have an arbitrary phase structure over $t$.  
For simplicity we will assume that $P$ and $Q$ have constant phase; and so without loss of generality we can simply take them as both being real and positive in the range $0 \le \{P(t),Q(t)\} \le 1$ everywhere, in keeping with the beam splitter model.

\smallskip
We this in hand we can now start from Equation~\ref{eq:HOM_term_recast} and arrive at
\begin{eqnarray}
    ((A \, B^{*}) \star (A \, B^{*}))(\Delta t) & = &
  (((XP) \, (YQ)^{*}) \, \star \, ((XP)\, (YQ)^{*} ))(\Delta t) 
     \nonumber \\
  & = & 
  (((XY^{*}) \, (PQ)^{*}) \, \star \, ((XY^{*}) \, (PQ)^{*}))(\Delta t) \nonumber \\
  & = &\mathcal{F}^{-1} \left[ \, \left| 
  (\mathcal{F} [ XY^{*} ] \star \mathcal{F} [ PQ ] )(\nu) \right|^{2} \right] \nonumber \\
  & = &
  \mathcal{F}^{-1} \left[ \, \left| 
  \left( (x \star y) \star \mathcal{F} [ PQ ] \right) (\nu) \right|^{2} \right] 
  \label{eq:HOM_term_recast_pulsed_1}
\end{eqnarray}
% 
%Here the last step makes use of the cross-correlation corollary of the {\em convolution theorem}.  The corollary states that the Fourier transform of the product of a function and the complex conjugate of another function is equal to the cross-correlation between the Fourier transforms of the two functions, e.g. exactly
%$\mathcal{F} [ XY^{*} ] =x \star y$, given the  definitions in Equation~\ref{eq:X_Y_x_y}.

Equation~\ref{eq:HOM_term_recast_pulsed_1} is so far fully general.  But now we can make use of the properties of $x(\nu)$ and $y(\nu)$, and the $(x \star y)(\nu)$ cross-correlation, as having essentially random phases over $\nu$, since they are narrow-band CW beams.  Following the logic leading from Eq.~\ref{eq:HOM_core_with_exp} to Eq.~\ref{eq:HOM_core_final}, now based on Eq.~\ref{eq:x_y_phase}, we can now simplify Equation~\ref{eq:HOM_term_recast_pulsed_1} in two steps to 
\begin{eqnarray}
    ((A \, B^{*}) \star (A \, B^{*}))(\Delta t) 
  & = &
  \mathcal{F}^{-1} \left[ \, \left| 
  \left( (x \star y) \star \mathcal{F} [ PQ ] \right )(\nu) \right|^{2} \right]
  \nonumber \\
  & = &
  \mathcal{F}^{-1} \left[ \, 
  \left( (|x \star y|^{2}) \star |\mathcal{F} [ PQ ]|^2 \right) (\nu) \right]
  \nonumber \\
  & = &
  \mathcal{F}^{-1} \left[ \,  
  \left( (|x|^2 \star |y|^{2}) \star |\mathcal{F} [ PQ ]|^2 \right) (\nu)  \right]
  \label{eq:HOM_term_recast_pulsed_complete}
\end{eqnarray}
Substituting back into the master Equation~\ref{eq:pair_rate_Dt_autocorr} for the pairs rate we have
\begin{equation}
    \frac{dN^{(2)}}{d \, \Delta t}  \; \propto \;
    \frac{1}{4} ((I_{a}+I_{b}) \star (I_{a}+I_{b}))(\Delta t) 
    - \frac{1}{2} \textbf{Re} \left\{ 
    \mathcal{F}^{-1} \left[ \, 
    \left( (|x(\nu)|^2 \star |y(\nu)|^2) \star |\mathcal{F} [ PQ ]|^2
    \right) (\nu) \,
    \right](\Delta t) \right\}
    \label{eq:pair_rate_Dt_pulsed_final}
\end{equation}
as the generalized version of Equation~\ref{eq:pair_rate_Dt_CW_final}, now for pulsed beams.
Equation~\ref{eq:pair_rate_Dt_pulsed_final} is now completely actionable, involving only three well-defined,  real-valued functions: the two parent CW beams' spectral line shapes $|x(\nu)|^{2}$ and $|y(\nu)|^2$ and the product $(PQ)(t)$ of the two pulse envelope amplitude functions.  

%%%%%%%%%%%%%%%%%%%%%%%%%%%%%%
% Examples with pulsed beams
%
%\subsubsection{Examples with pulsed beams}
%\label{subsec:pulsed_examples}

\paragraph{Single, synchronous, symmetric pulses. }\label{subsubsec:single_synchronous_Gaussians}We look first at the single envelop case, namely where (i)~the beam envelopes contain only one well-shaped pulse each; (ii)~the pulses both arrive at the beam splitter simultaneously; and (iii)~the shapes of the pulses are symmetric around their peaks; this lets us conveniently set $t=0$ as the peak arrival time for both $P(t)$ and $Q(t)$.  In this case the product transform $\mathcal{F} [ PQ ](\nu)$ will be real and symmetric around $\nu=0$, and the cross-correlations will be easy.

We also assume the parent beam lineshapes and the pulse amplitude shapes are all Gaussians.  For further simplicity we assign the parent beams to have power levels $I_{a}^{Max}$ and $I_{b}^{Max}$ and assume $P(0)=Q(0)=1$ at the pulse peak.  Then, following on Equation~\ref{eq:alpha_beta_Gaussians} we set
\begin{eqnarray}
    |x|^2(\nu) & = & \frac{I_{a}^{Max}}{\sigma_{\nu\;beam} \sqrt{2 \pi}} \; g(\nu; \nu_{0}+\delta\nu, \, \sigma_{\nu\;beam}) 
    \nonumber \\
    |y|^2(\nu) & = & \frac{I_{b}^{Max}}{\sigma_{\nu\;beam} \sqrt{2 \pi}} \; g(\nu; \nu_{0}, \, \sigma_{\nu\;beam}) 
    \label{eq:x_y_Gaussians} \\
    P(t) &= & Q(t) = G(t; 0, \sqrt{2} \, \sigma_{t\;pulse})
    \label{eq:P_Q_Gaussians} \\
    I_{a}(t) & = & I_{a}^{Max} (P(t))^{2} 
    = I_{a}^{Max} G(t; 0, \sigma_{t\;pulse}) 
    \nonumber \\    
    I_{b}(t) & = & I_{b}^{Max} (Q(t))^{2}
    = I_{b}^{Max} G(t; 0, \sigma_{t\;pulse})
    \label{eq:I_a_I_b_Gaussians}
\end{eqnarray}
where we have assumed that the two parent beam lineshapes have the same Gaussian frequency standard deviation $\sigma_{\nu\;beam}$ and the two pulses have the same time standard deviation $\sigma_{t\;pulse}$; and we allow for a small difference in central frequency $\delta \nu$ between the parent beams but have the two pulses perfectly synchronized.  Note that we are choosing the convention that $\sigma_{t\;pulse}$ describes the Gaussian width of the pulse in the intensity $I(t)$, and so the Gaussian widths of the amplitude modulations $P(t)$ and $Q(t)$ are greater by a factor of $\sqrt{2}$.

\smallskip
We first evaluate the classical term from Eq.~\ref{eq:pair_rate_Dt_pulsed_final}:
\begin{eqnarray}
    \frac{1}{4} ((I_{a}+I_{b}) \star (I_{a}+I_{b}))(\Delta t) 
    & = & \frac{1}{4}\left( (I_{a}^{Max} + I_{b}^{Max})^2
    \; (G(t; 0, \sigma_{t\;pulse}) \star 
      G(t; 0, \sigma_{t\;pulse}))(\Delta t)   
    \right) \nonumber \\
    & = & \frac{1}{4} (I_{a}^{Max} + I_{b}^{Max})^2
    \;  \sqrt{\pi} \; \sigma_{t\;pulse}  \;
    G(\Delta t; 0, \sqrt{2} \, \sigma_{t\;pulse}). 
    \label{eq:rate_dt_pulsed_Gaussian_classical_term}
\end{eqnarray}
%This is just a standard cross-correlation result, that the distribution of the pair time difference is $\sqrt{2}$ times wider than the time distribution of each single distribution.

The interference term is also quite straightforward for the simple case; first for the pulse shape factor:
\begin{eqnarray}
    \left| \mathcal{F} [ PQ ] \right|^2 & = &
    \left| \mathcal{F} [ G(t; 0,\sigma_{t\;pulse}) ] \right|^2 \nonumber \\
    & = &  
    \left|  \sigma_{t\;pulse} \sqrt{2 \pi}  \;
    g(\nu; 0, 1/2 \pi \sigma_{t\;pulse}) \right|^2 
    \nonumber \\
    & = &  
    2 \pi \, {\sigma_{t\;pulse}}^2  \;
    g(\nu; 0, 1/ 2 \sqrt{2} \, \pi \, \sigma_{t\;pulse}) 
    \nonumber
\end{eqnarray}
And for the parent CW beam width factor:
\begin{eqnarray}
    |x(\nu)|^2 \star |y(\nu)|^2
    & = &
    \frac{I_{a}^{Max} \, I_{b}^{Max}}
      {{\sigma_{\nu\;beam}}^2 \; 2\pi} \;
    \left( g(\nu; \nu_{0}+\delta\nu, \, \sigma_{\nu\;beam}) 
      \star
    g(\nu; \nu_{0}, \, \sigma_{\nu\;beam}) \right)(\nu)
    \nonumber \\
     & = &
    \frac{I_{a}^{Max} \, I_{b}^{Max}}
      {{\sigma_{\nu\;beam}}^2 \; 2\pi} \;
      \sqrt{\pi} \, \sigma_{\nu\;beam} \;
      g(\nu; \delta \nu, \, \sqrt{2} \; \sigma_{\nu\;beam})
\end{eqnarray}
Subsituting into Equation~\ref{eq:pair_rate_Dt_pulsed_final}, carrying out the final cross-correlation, inverse Fourier transform, and taking the real part we arrive at the full interference term:
\begin{eqnarray}
 & & 
- \frac{1}{2} \textbf{Re} \left\{ 
    \mathcal{F}^{-1} \left[ \, 
    \left( (|x|^2 \star |y|^2) \star |\mathcal{F} [ PQ ]|^2
    \right)  \,
    \right] \right\}  =  \nonumber \\
    & & \quad \quad
    -\frac{I_{a}^{Max} \, I_{b}^{Max}}{2} \sqrt{\pi} \; \sigma_{t\;pulse} \,
    G\left( \Delta t;  0, 
    \frac{\sqrt{2} \; \sigma_{t\;pulse}}{\sqrt{1+(4 \pi \sigma_{\nu\;beam} \, \sigma_{t\;pulse}})^2 }
    \right) \, \cos{(2 \pi \, \delta \nu \, \Delta t )}
    \label{eq:rate_dt_pulsed_Gaussian_quantum_term}
\end{eqnarray}
Adding the two terms from Eq.~\ref{eq:rate_dt_pulsed_Gaussian_classical_term} and Eq.~\ref{eq:rate_dt_pulsed_Gaussian_quantum_term}, and after some simplification, we can write the pairs rate in the case of Gaussian beams and pulses as
\begin{eqnarray}
   \frac{dN^{(2)}}{d \, \Delta t}  \; & \propto & \; 
    G(\Delta t; 0, \sqrt{2} \, \sigma_{t\;pulse}) 
    -\mathcal{V} \;
    G\left( \Delta t;  0, 
    \frac{\sqrt{2} \; \sigma_{t\;pulse}}{\sqrt{1+(4 \pi \sigma_{\nu\;beam} \, \sigma_{t\;pulse}})^2 }
    \right) \, \cos{(2 \pi \, \delta \nu \, \Delta t )}
    \label{eq:pair_rate_Dt_pulsed_Gaussian}  \\
    \mathrm{with} \quad
    \mathcal{V}  & \equiv & 
    \frac{2 I_{a}^{Max} I_{b}^{Max}}{(I_{a}^{Max}+I_{b}^{Max})^2}
    \nonumber
\end{eqnarray}
The visibility $\mathcal{V}$ is defined from the relative intensities exactly as in Equation~\ref{eq:pair_rate_Dt_CW_Gaussian_visnorm}, and reaches a maximum value of 0.5 when the beam peak intensities are equal.

\smallskip
Equation~\ref{eq:pair_rate_Dt_pulsed_Gaussian} is now a complete form for the pairs rate in the case of single, synchronized Gaussian pulses, with only four shape parameters: pulse width, parent beam lineshape width, parent beam frequency offset, and HOM visibility at $\Delta t = 0$.  We can make quick checks on its behavior in the two natural limits, being very wide pulses and very narrow pulses.

\begin{itemize}
    \item In the limit that $\sigma_{t\;pulse}$ is large the first Gaussian of Eq.~\ref{eq:pair_rate_Dt_pulsed_Gaussian} approaches a constant with value 1 in the neighborhood around $\Delta t = 0$.  At the same time the standard deviation of the second Gaussian will approach $\sigma=1/\sqrt{2} 2 \pi \sigma_{\nu\;beam}$; and so Eq.~\ref{eq:pair_rate_Dt_pulsed_Gaussian} will reduce to exactly the CW case result of Equation~\ref{eq:pair_rate_Dt_CW_Gaussian_visnorm} as we would expect.

    \item In the limit that $\sigma_{t\;pulse}$ is small, specifically small compared to $1/2\pi \sigma_{\nu\;beam}$, we can see that the sigma of the second Gaussian will approach $\sigma = \sqrt{2} \, \sigma_{t\;pulse}$, i.e. the same as that of the first Gaussian.  We can then factor out the common Gaussian to arrive at
    \begin{equation}
 \frac{dN^{(2)}}{d \, \Delta t}  \;  \propto  \; 
    \left( 1- \mathcal{V} \cos{(2 \pi \, \delta \nu \, \Delta t )}
    \right) \; G(\Delta t; 0, \sqrt{2} \, \sigma_{t\;pulse})   
    \end{equation}
    Thus in the limit of short pulses, i.e. much shorter than the parent beams' coherence time, we have the extremely simple result that the HOM interference has the effect of reducing the height of the peak in the pairs distribution at $\Delta t = 0$ by a factor of $(1-\mathcal{V})$, but leaving its shape the same, if the frequency mismatch $\delta \nu$ is small enough that the cosine factor is essentially constant.
\end{itemize}

%%%%%%%%%%%%%%%%%%%%%%%%%%%
%
% Extenstion from single pulse to pulse train; overall simple
%
\paragraph{Extension to a regular train of separated, synchronous, symmetric pulses. }\label{subsubsec:extension_to_Gaussian_pulsetrain}In the experiment we create not just a single pulse at each input but a pulse train.  We can extend the above analysis to pulse train input beams, provided we satisfy the criterion that the pulses are well-separated.  We  make two assumptions:

\begin{itemize}
    \item Each pulse is over by the time the next one starts, e.g. at any given time only one pulse in the train can have non-negligible intensity.

    \item The interval $\Delta T$ between pulses is long compared to the widths of the pulses themselves.   
\end{itemize}

% As a rough rule of thumb in the case of Gaussians we would want $\Delta T \ge 10 \, \sigma_{t\;pulse}$, which will be sure to satisfy both these conditions.  If the pulses are not well-separated in the above sense then one can certainly still calculate an HOM interference effect, but we would have to consider an harmonic analysis of the pulse train function rather than examining the profile of each individual pulse.

Qualitatively speaking, in the case that the pulses in the train are well-separated, as above, and the arrival of a pulse in one input channel at the beam splitter is always synchronized with a pulse in the other channel, then we can think of the experiment as a repetition at intervals of the single-pulse version. The analysis of the ``central'' peak around $\Delta t=0$ in the pairs distribution remains exactly as in the single-pulse case.  The only change in the pairs distribution is the appearance of pairs with larger $\Delta t$'s, away from $\Delta t=0$, which correspond to the arrival of photons pairs from non-synchronous pulses. In this case of well-separated pulses the ``off-beat'' pairs will not show any detectable effect of quantum interference: only the central pairs peak shows an effect, and that is identical to that in the single-pulse case.

\smallskip
With this in hand we can now write the result for a regular train of synchronized Gaussian pulses at the input channels.  All we need to do is to copy over Equation~\ref{eq:pair_rate_Dt_pulsed_Gaussian} and repeat the classical term at intervals with spacing $\Delta T$:

\begin{equation}
\begin{aligned}
   \frac{dN^{(2)}}{d \, \Delta t}  \;  &\propto  \; 
   \left[ \sum_{n}
    G(\Delta t; n \, \Delta T, \sqrt{2} \, \sigma_{t\;pulse})   
    \right]\\
    &-\mathcal{V} \;
    G\left( \Delta t;  0, 
    \frac{\sqrt{2} \; \sigma_{t\;pulse}}{\sqrt{1+(4 \pi \sigma_{\nu\;beam} \, \sigma_{t\;pulse}})^2 }
    \right) \, \cos{(2 \pi \, \delta \nu \, \Delta t )}.
    \label{eq:pair_rate_Dt_pulsetrain_Gaussian}  
\end{aligned}
\end{equation}
\clearpage

\end{document}